\documentclass{aastex63}
\usepackage{CJKutf8}
\usepackage[caption=false]{subfig}
\usepackage{graphicx}
\usepackage{epstopdf}
\usepackage{longtable}
\usepackage{multirow}
\usepackage{graphicx}
\usepackage{booktabs}
\usepackage{amsmath,amssymb}
\usepackage[T1]{fontenc}
\graphicspath{{./}{figures/}}
\newcommand{\ra}[4]{$#1^{\rm h}#2^{\rm m}#3^{\rm s}.#4$}
\newcommand{\dec}[4]{$#1^{\circ}#2'#3''.#4$}

\begin{document}

\title{The Study of Dust Formation of Six Tidal Disruption Events}

\correspondingauthor{Shan-Qin Wang \begin{CJK*}{UTF8}{gbsn}(王善钦)\end{CJK*},Wen-Pei Gan \begin{CJK*}{UTF8}{gbsn}(甘文沛)\end{CJK*}}
\email{shanqinwang@gxu.edu.cn, wpgan@niaot.ac.cn}

\author{Xian-Mao Cao \begin{CJK*}{UTF8}{gbsn}(曹先茂)\end{CJK*}}
\affiliation{Guangxi Key Laboratory for Relativistic Astrophysics,
School of Physical Science and Technology, Guangxi University, Nanning 530004,
China}

\author{Shan-Qin Wang \begin{CJK*}{UTF8}{gbsn}(王善钦)\end{CJK*}}
\affiliation{Guangxi Key Laboratory for Relativistic Astrophysics,
School of Physical Science and Technology, Guangxi University, Nanning 530004,
China}

\author{Wen-Pei Gan \begin{CJK*}{UTF8}{gbsn}(甘文沛)\end{CJK*}}
\affiliation{Guangxi Key Laboratory for Relativistic Astrophysics,
School of Physical Science and Technology, Guangxi University, Nanning 530004, China}
\affiliation{Nanjing Institute of Astronomical Optics \& Technology, Nanjing 210042, China}

\author{Jing-Yao Li  \begin{CJK*}{UTF8}{gbsn}(李京谣)\end{CJK*}}
\affiliation{Guangxi Key Laboratory for Relativistic Astrophysics,
School of Physical Science and Technology, Guangxi University, Nanning 530004,
China}

\begin{abstract}

This paper investigates eleven (UV-)optical-infrared
(IR) spectral energy distributions (SEDs) of six tidal disruption events (TDEs),
which are ASASSN-14li, ASASSN-15lh, ASASSN-18ul, ASASSN-18zj, PS18kh, and ZTF18acaqdaa.
We find that all the SEDs show evident IR excesses.
We invoke the blackbody plus dust emission model to fit the SEDs,
and find that the model can account for the SEDs.
The derived masses of the dust surrounding ASASSN-14li, ASASSN-15lh,
ASASSN-18ul, ASASSN-18zj, PS18kh, and ZTF18acaqdaa are respectively
$\sim0.7-1.0\,(1.5-2.2)\times10^{-4}\,M_\odot$,
$\sim0.6-3.1\,(1.4-6.3)\times10^{-2}\,M_\odot$,
$\sim1.0\,(2.8)\times10^{-4}\,M_\odot$,
$\sim0.1-1.6\,(0.3-3.3)\times10^{-3}\,M_\odot$,
$\sim1.0\,(2.0)\times10^{-3}\,M_\odot$, and
$\sim 1.1\,(2.9)\times10^{-3}\,M_\odot$, if the dust is graphite (silicate).
The temperature of the graphite (silicate) dust of the six TDEs are respectively
$\sim1140-1430\,(1210-1520)$\,K, $\sim1030-1380\,(1100-1460)$\,K, $\sim1530\,(1540)$\,K,
$\sim960-1380\,(1020-1420)$\,K, $\sim900\,(950)$\,K, and $\sim1600\,(1610)$\,K.
By comparing the derived temperatures to the vaporization temperature
of graphite ($\sim 1900$\,K) and silicate ($\sim 1100-1500$\,K),
we suggest that the IR excesses of PS18kh can be explained by both the
graphite and silicate dust, the rest five TDEs favor the
graphite dust while the silicate dust model cannot be excluded.
Moreover, we demonstrate the lower limits of the radii of the dust shells
surrounding the six TDEs are significantly larger than those of the radii of the
photospheres at the first epochs of SEDs, indicating that the dust might
exist before the the TDEs occurred.

\end{abstract}


\section{Introduction}
\label{sec:intro}

Supermassive black holes in the centers of galaxies can disrupt stars,
and produce the tidal disruption events (TDEs, \citealt{Hills1975,Rees1988,van Velzen2016,Gezari2021}),
which emit luminous multi-band radiation \citep{Frederick2019}.
Generally, the X-ray and optical light curves of TDEs follow a power-law function ($F_\nu \propto t^{-5/3}$)
which reflects the evolution of the accretion rate. The spectral energy distributions (SEDs) of most
TDEs can be well described by the blackbody function, while the SEDs of some TDEs cannot
be fitted by the blackbody function and show evident infrared (IR) excesses
relative to the blackbody SEDs (e.g., \citealt{Dou2016,Jiang2016,Dou2017,Jiang2017,Jiang2019}).

It is widely believed that the IR excesses of SEDs of TDEs, SNe, as well
as other optical transients can be explained by the dust emission.
In this scenario, the UV and optical emission from the photospheres
of the transients heat the dust surrounding them, and the heated dust
emit flux peaked at IR bands and produce IR excesses. The flux from the
photospheres and the dust shells yield SEDs with IR excesses relative
the blackbody SEDs.
Studying the IR excesses of the SEDs of TDEs would help to infer the masses,
temperature, chemical composition, and locations of the dust grains.

Based on the observations of $W_1$ (3.4 $\mu${m}) and $W_2$ (4.6 $\mu${m})
of {\it Wide-field Infrared Survey Explorer} ($WISE$) and the optical
data obtained by optical telescopes, \cite{Jiang2016} construct 3
optical-IR SEDs of ASASSN-14li, and use double-blackbody model to
derive temperatures of the dust at different epochs by fitting the SEDs.
The derived value dust temperature at MJD $\approx$ 57019 is 2091$\pm$172\,K.
\cite{Jiang2017} use double-blackbody model and blackbody plus dust
model to fit the optical-IR SED of PS16dtm at MJD $\approx$ 57,755, finding that
the dust temperature is 1180 K (for the former model) or 700\,K
(for the later model).

\cite{Jiang2021} collect a sample of TDEs observed by
$WISE$ at the filters of $W_1$ and $W_2$. To subtract the flux from the
photospheres of the TDEs, they adopt the evolution pattern of the photosphere bolometric
luminosity obtained by \cite{van Velzen2020}, and assume that the temperature of the
TDEs keep constant to be the averaged values of the first 100 days.
After subtracting the flux from the photospheres in $W_1-$ and $W_2-$ bands,
\cite{Jiang2021} get the IR excesses at all epochs for all $WISE$-selected
TDEs. They invoke the dust model to fit the IR excesses of the SEDs of
TDEs and derive the physical parameters (temperature, luminosity, radii of the dust shell,
covering factor, etc.) of the dust.

In this paper, we combine the IR ($W_1$, $W_2$) photometry of six TDEs
provided by \cite{Jiang2021} with the (UV-)optical photometry of the same TDEs at
the same epochs, constructing the (UV-)optical-IR SEDs and fit the SEDs by
using the blackbody model to test if the SEDs of the six TDEs show IR excesses.
For the SEDs showing IR excesses, we use blackbody plus dust emission model to fit them,
and constrain the physical properties of the dust.
This method does not depend on any assumptions about the
bolometric luminosity and the temperature evolution.

The paper is organized as follow. In section \ref{sec:SED},
we use the single-component model and double-component model to fit the SEDs of the
six TDEs. We discuss our results and draw some conclusions in
Sections \ref{sec:discussion} and \ref{sec:Con}, respectively.

\section{The Optical and IR SEDs and the SED fitting}
\label{sec:SED}

In this section, we study the SEDs of the six TDEs (ASASSN-14li, ASASSN-15lh,
ASASSN-18ul, ASASSN-18zj, PS18kh, and ZTF18acaqdaa) having opcital-IR SEDs.
The information of the six TDEs is listed in Table \ref{table:details}.
The IR photometry of all the six TDEs are from Table 2 of \cite{Jiang2021}.
The sources of the (UV-)optical photometry of the TDEs are also listed
in Table \ref{table:details}.

Combining the IR ($W_1$, $W_2$) photometry of \cite{Jiang2021} with the (UV-)optical
photometry, we obtain eleven SEDs of ASASSN-14li (3 SEDs), ASASSN-15lh
(3 SEDs) \footnote{It should be noted that the nature of ASASSN-15lh is still
in debate. \cite{Dong2016} suggest that it is a superluminous SN, while \cite{Leloudas2016}
argue that it is a TDE.}, ASASSN-18ul (1 SED), ASASSN-18zj
(2 SEDs), PS18kh (1 SED), and ZTF18acaqdaa (1 SED).


First, we fit the SEDs of the TDEs
using the blackbody model, in which
the flux from the photosphere ($F_{\nu,{\rm ph}}$) is
${\pi}B_\nu(T_{\rm ph}){R_{\rm ph}^2}/{D_L^2}$,
where $B_\nu(T_{\rm ph}) = (2 h{\nu}^3/c^2)(e^{\frac{h{\nu}}{k_{\rm b}T_{\rm ph}}}-1)^{-1}$
is the intensity of the blackbody radiation of the photosphere,
$R_{\rm ph}$, $T_{\rm ph}$, and $D_L$ are the radius, temperature of the
photosphere, and the luminosity distance of a TDE, respectively.
The free parameters of the model are the radius ($R_{\rm ph}$) and the
temperature ($T_{\rm ph}$) of the TDE photosphere.
The Markov Chain Monte Carlo (MCMC) method
using the \texttt{emcee} Python package \citep{Foreman-Mackey2013}
was adopted to get best-fitting parameters and the 1-$\sigma$ range of the parameters.

\begin{figure}[tbph]
\begin{center}
\includegraphics[width=0.25\textwidth,angle=0]{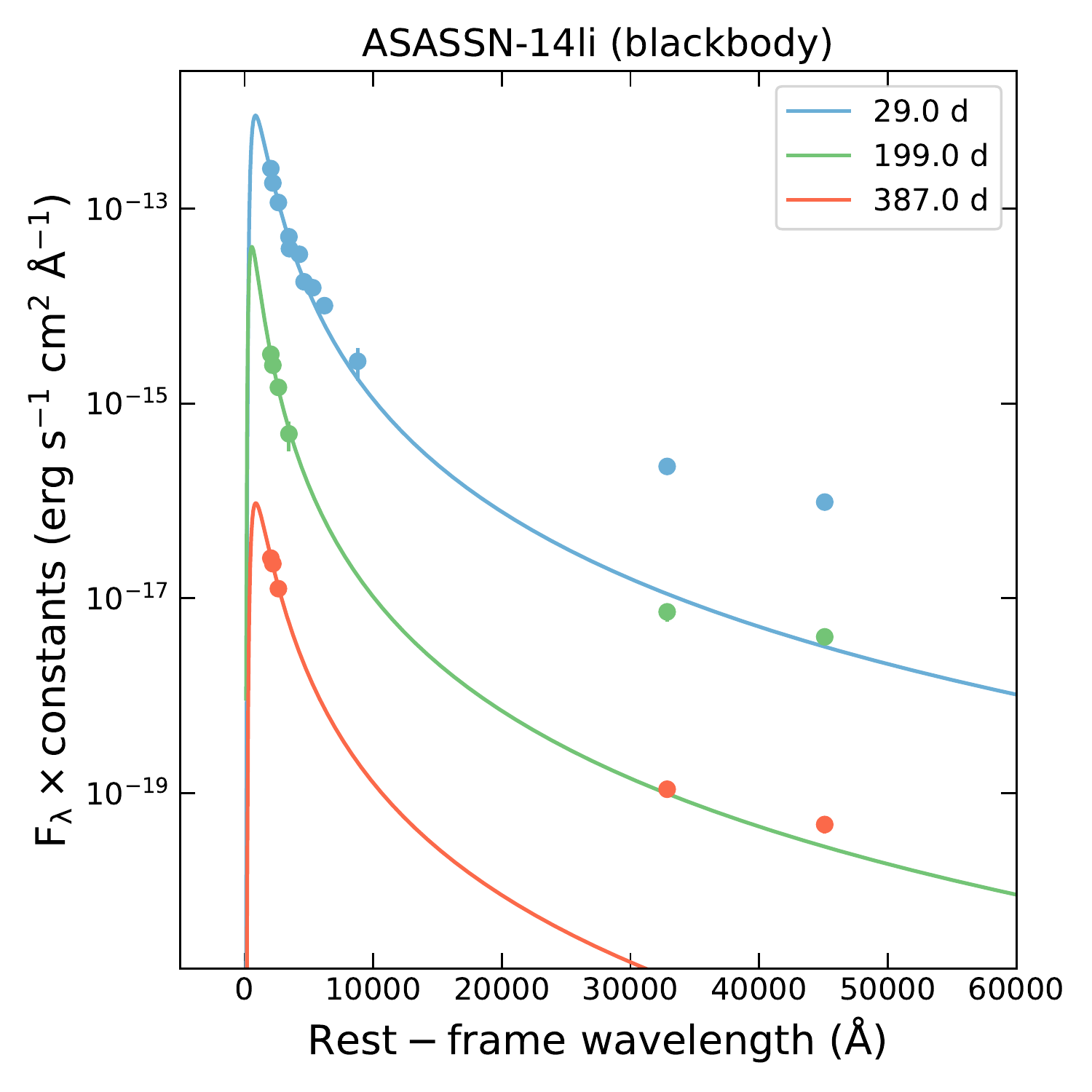}
\includegraphics[width=0.25\textwidth,angle=0]{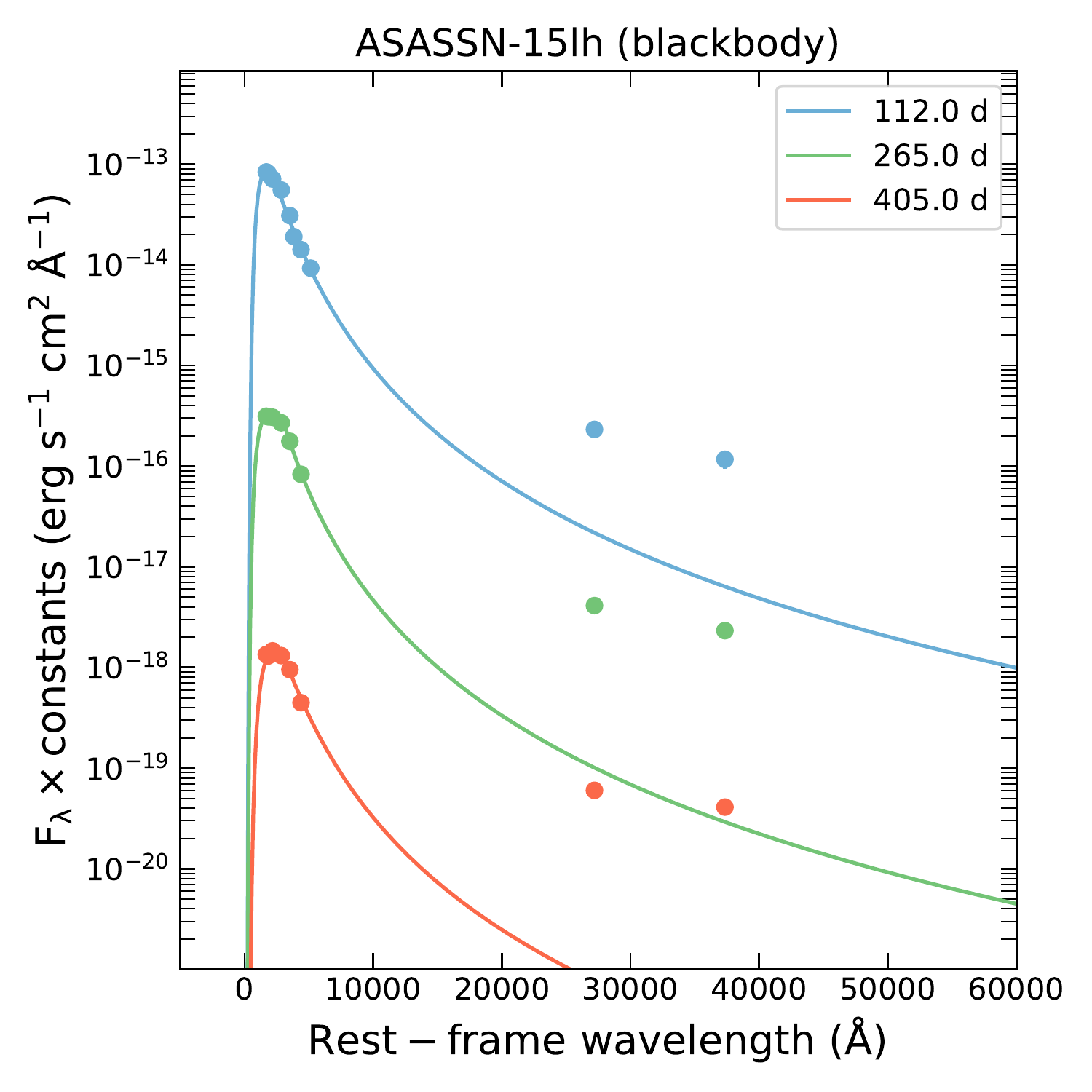}
\includegraphics[width=0.25\textwidth,angle=0]{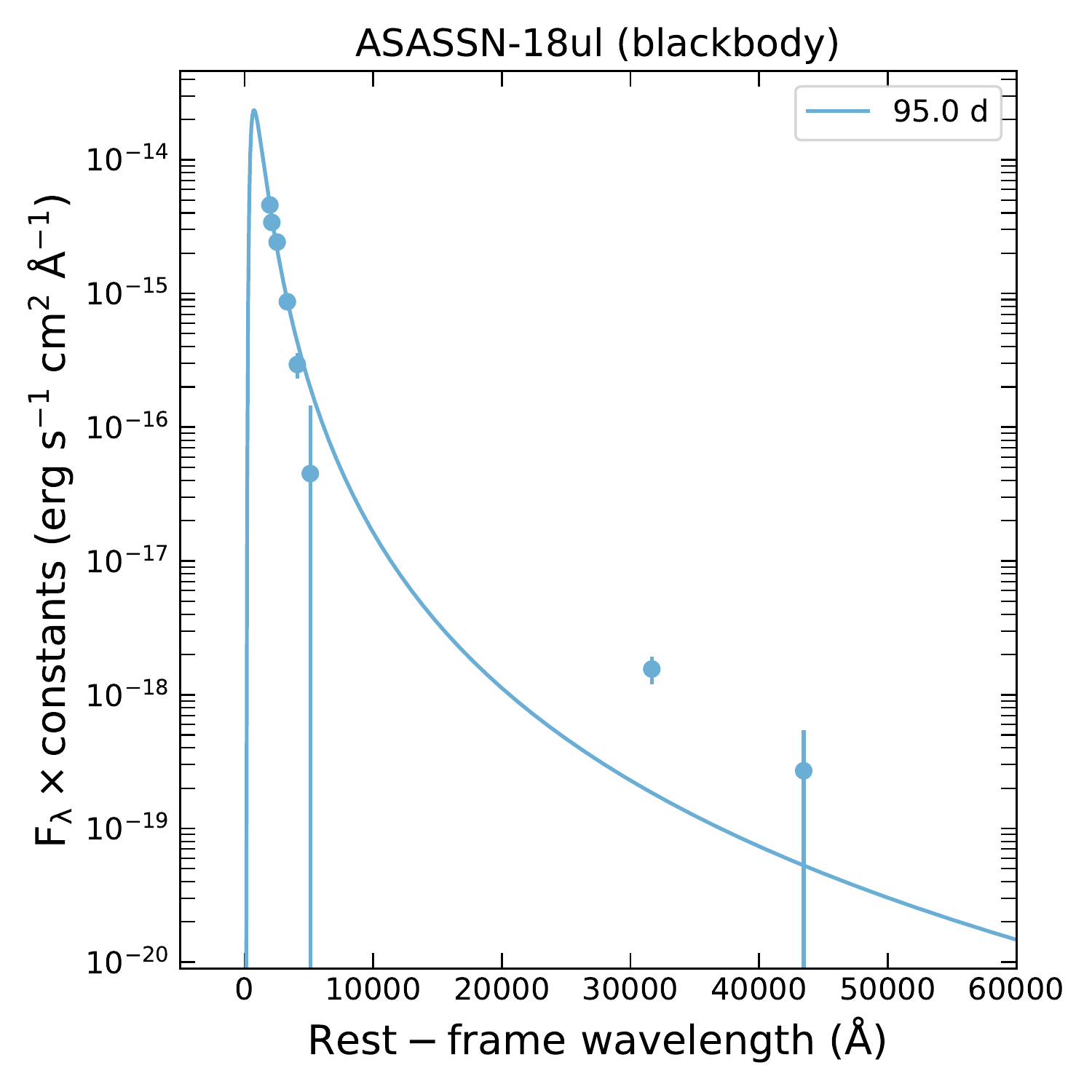}
\includegraphics[width=0.25\textwidth,angle=0]{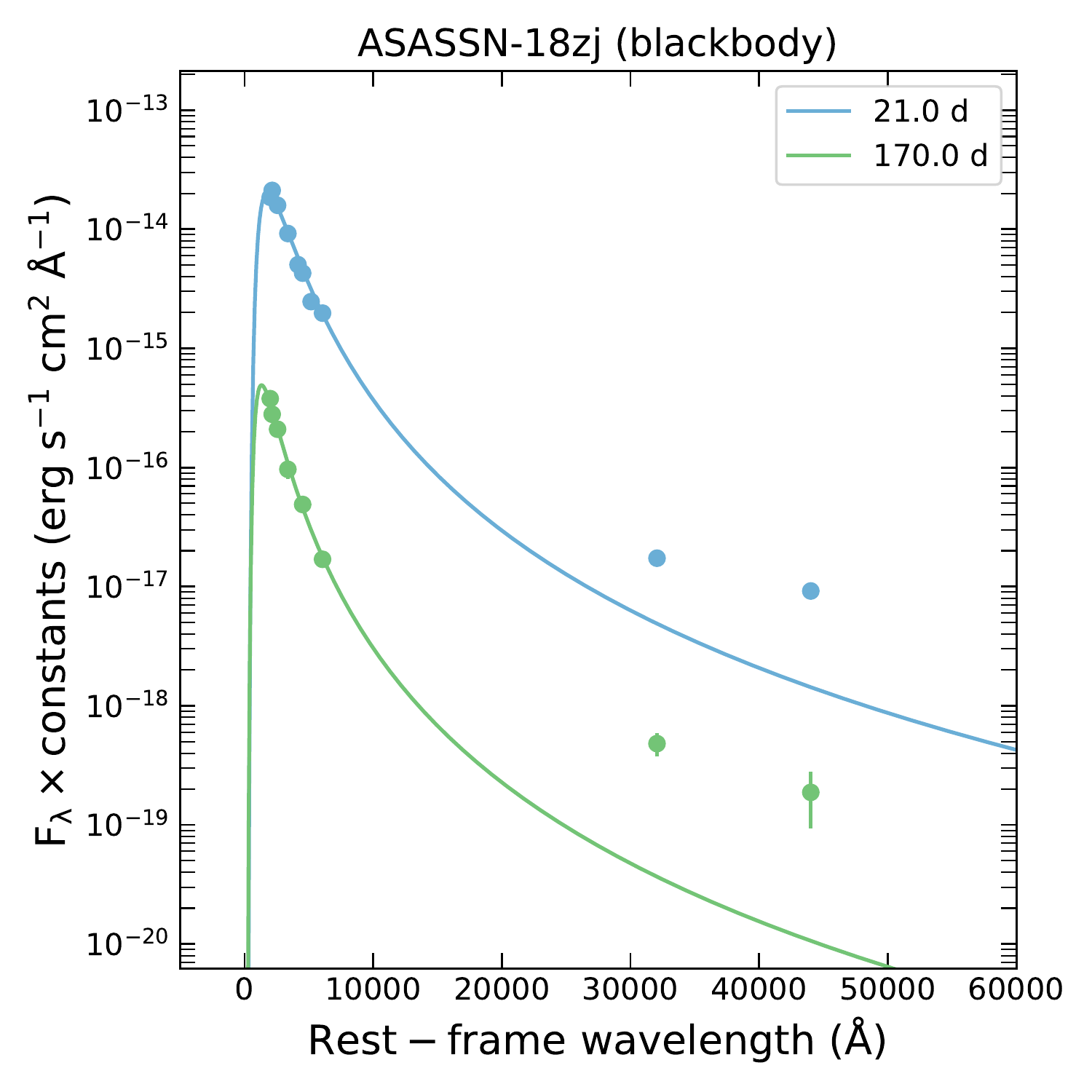}
\includegraphics[width=0.25\textwidth,angle=0]{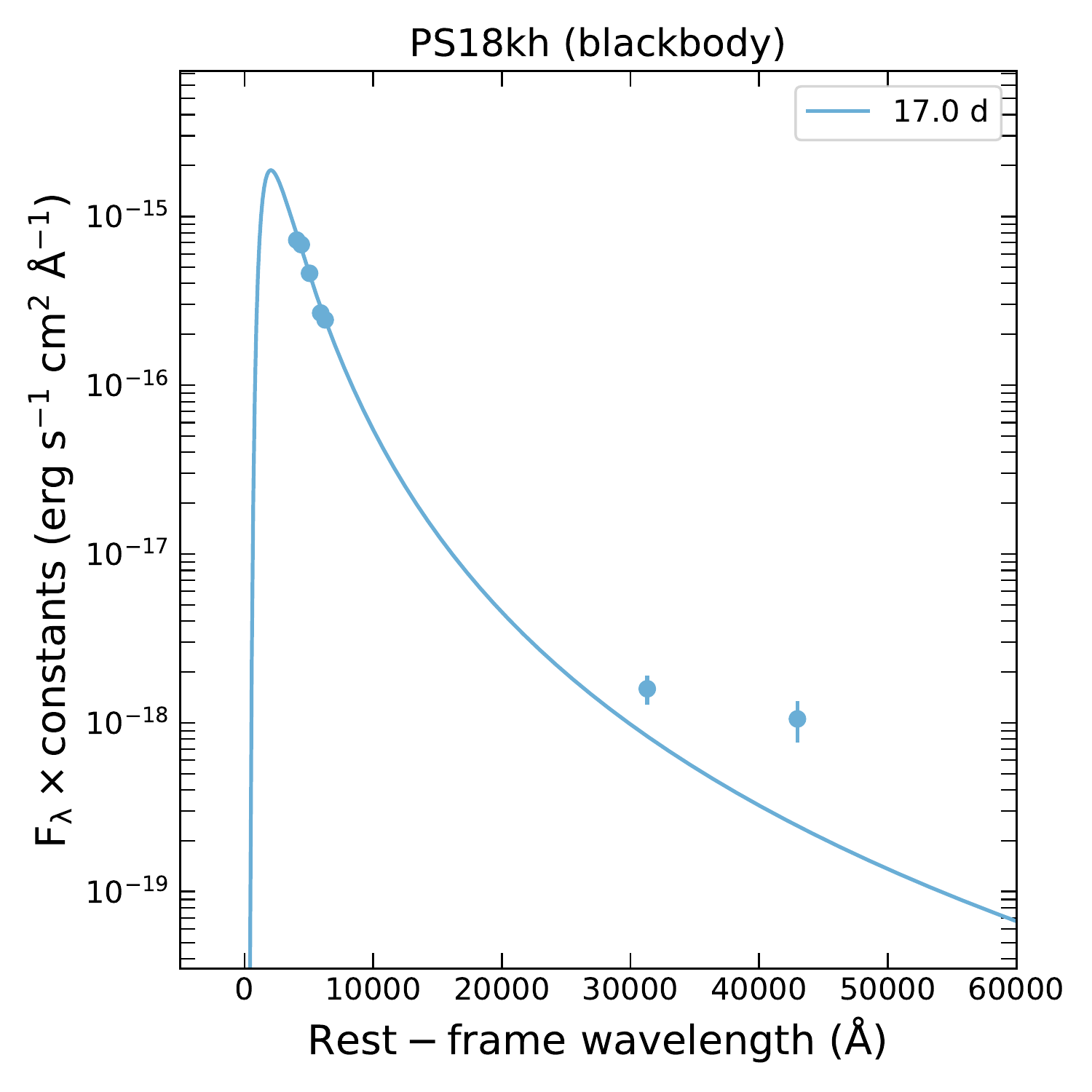}
\includegraphics[width=0.25\textwidth,angle=0]{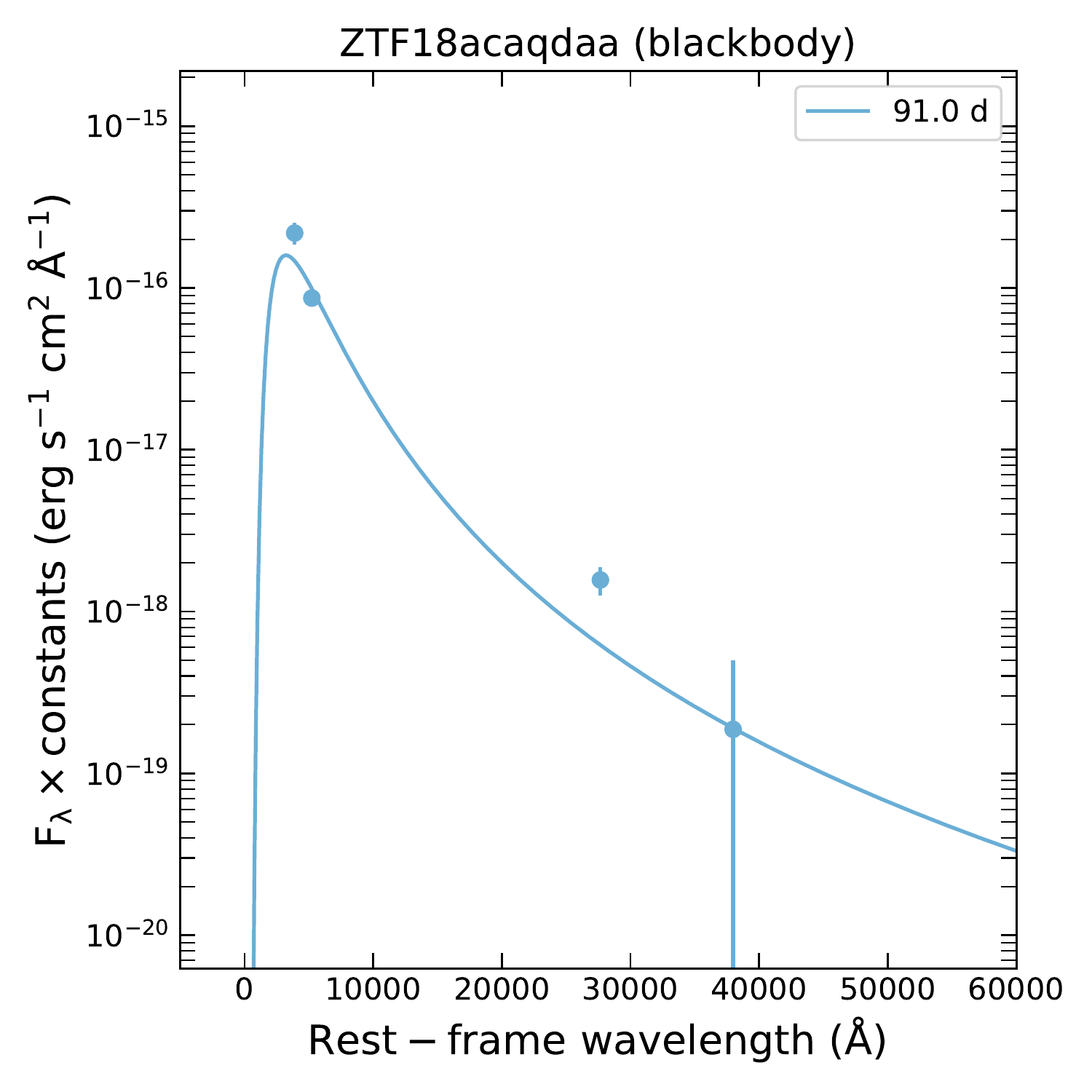}
\end{center}
\caption{The optical and IR SEDs of ASASSN-14li, ASASSN-15lh, PS18kh, ZTF18acaqdaa, ASASSN-18ul, and ASASSN-18zj
and the fits of the blackbody model. For clarity, the flux at all epochs are shifted
by multiplying different constants.}
\label{fig:SED1}
\end{figure}

The blackbody fits for the SEDs of the six TDEs and the
corresponding best-fitting parameters of the model are
shown in Figure \ref{fig:SED1} and Table \ref{table:SED_PARAM-single}, respectively.
Figure \ref{fig:SED1} indicates that all the SEDs of the six TDEs show evident IR
excesses relative to the blackbody SEDs.

To fit the SEDs, we assume that the IR excesses are
due to the emission from the dust heated by the UV and optical emission of the TDE
photospheres, and invoke the blackbody plus dust emission model to fit the SEDs.
We suppose that the dust can be either graphite or silicate, and
adopt a power-law grain size distribution function $f(a)\propto a^{-\alpha}$
(${\int }_{a_{\rm min} }^{a_{\rm max}}f\left(a\right)da=1$) \citealt{Gall2014}).
To compare the results with those of \cite{Jiang2021}, the
values of $\alpha$, $a_{\rm min}$, and $a_{\rm max}$ in this paper
are set to be 3.5, 0.01 $\mu$m, and 10 $\mu$m, respectively.

Assuming that the number of the dust grains having a radius of $a$ is $N_{d}(a)$,
the total mass of the dust $M_{d}={\int}_{a_{\rm min}}^{a_{\rm max}}N_{d}(a)m_{d}(a)da=N_{d}{\int}_{a_{\rm min}}^{a_{\rm max}}f(a)m_{d}(a)da$
($m_{d}$ is the the mass of a dust grain with a radius of $a$),
the flux from the dust is ${\int}_{a_{\rm min}}^{a_{\rm max}}[{\pi}{B}_{\nu}(T_{\rm d}){Q}_{\nu}(a)]4{\pi}a^2N(a)da/4{\pi}D_L^2$=
$N_{d}/D_{L}^{2}{\int}_{a_{\rm min}}^{a_{\rm max}}f(a)m(a){\kappa}_{\nu}(a){B}_{\nu}(T_{\rm d})da$
(since $\kappa_\nu (a) = \frac{3Q_\nu(a)}{4 \rho a}$, $m(a)=4/3\pi{\rho}a^3$).
Here, $T_{\rm d}$ is the temperature of the dust; ${\kappa}_{\nu}\left(a\right)$ is the mass
absorption coefficient of the dust grains (see \citealt{Gan2021} and the references therein for more details).
The free parameters of the blackbody plus dust emission model are
$R_{\rm ph}$, $T_{\rm ph}$, $T_{\rm d}$, and $M_{d}$.

\begin{figure}[tbph]
\begin{center}
\includegraphics[width=0.25\textwidth,angle=0]{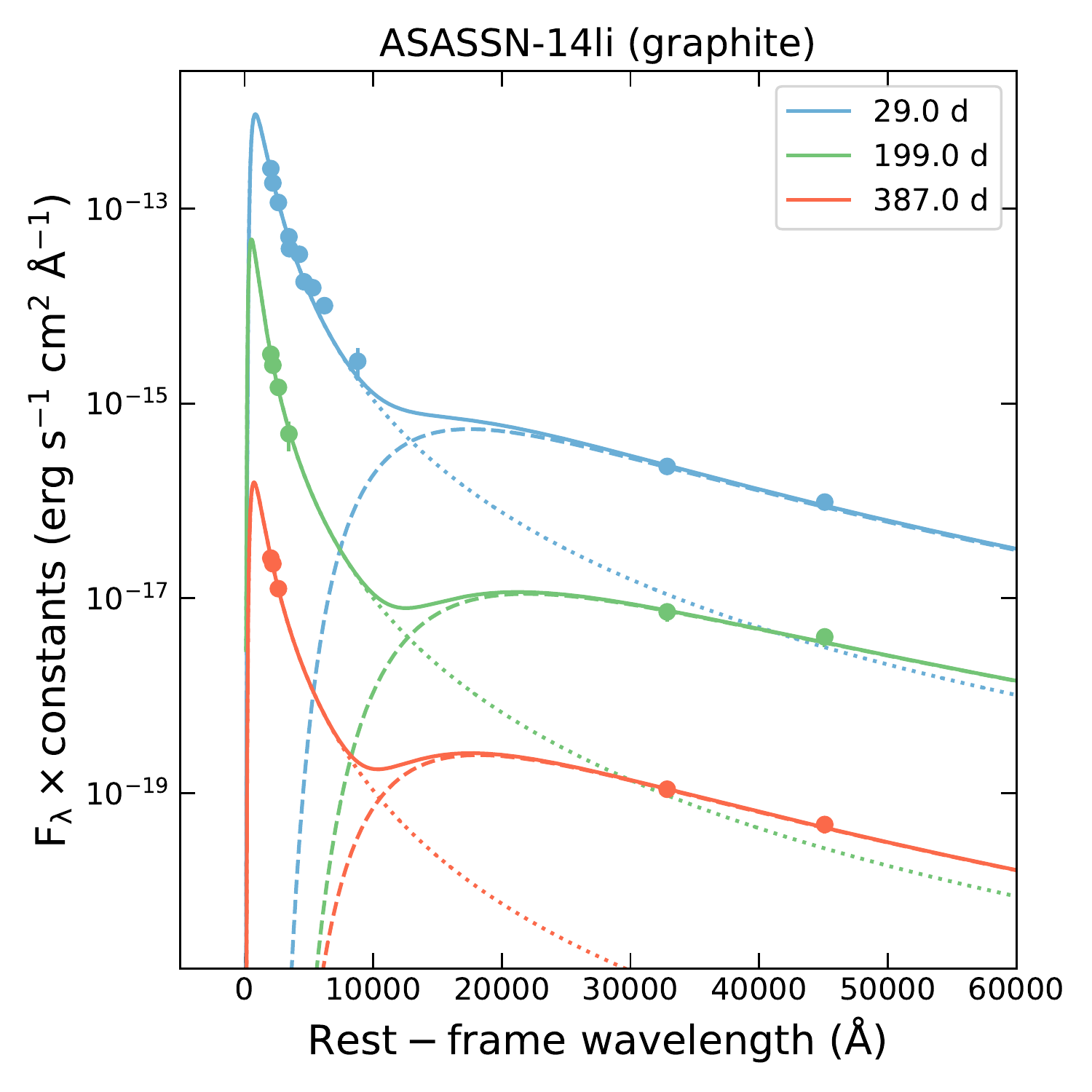}
\includegraphics[width=0.25\textwidth,angle=0]{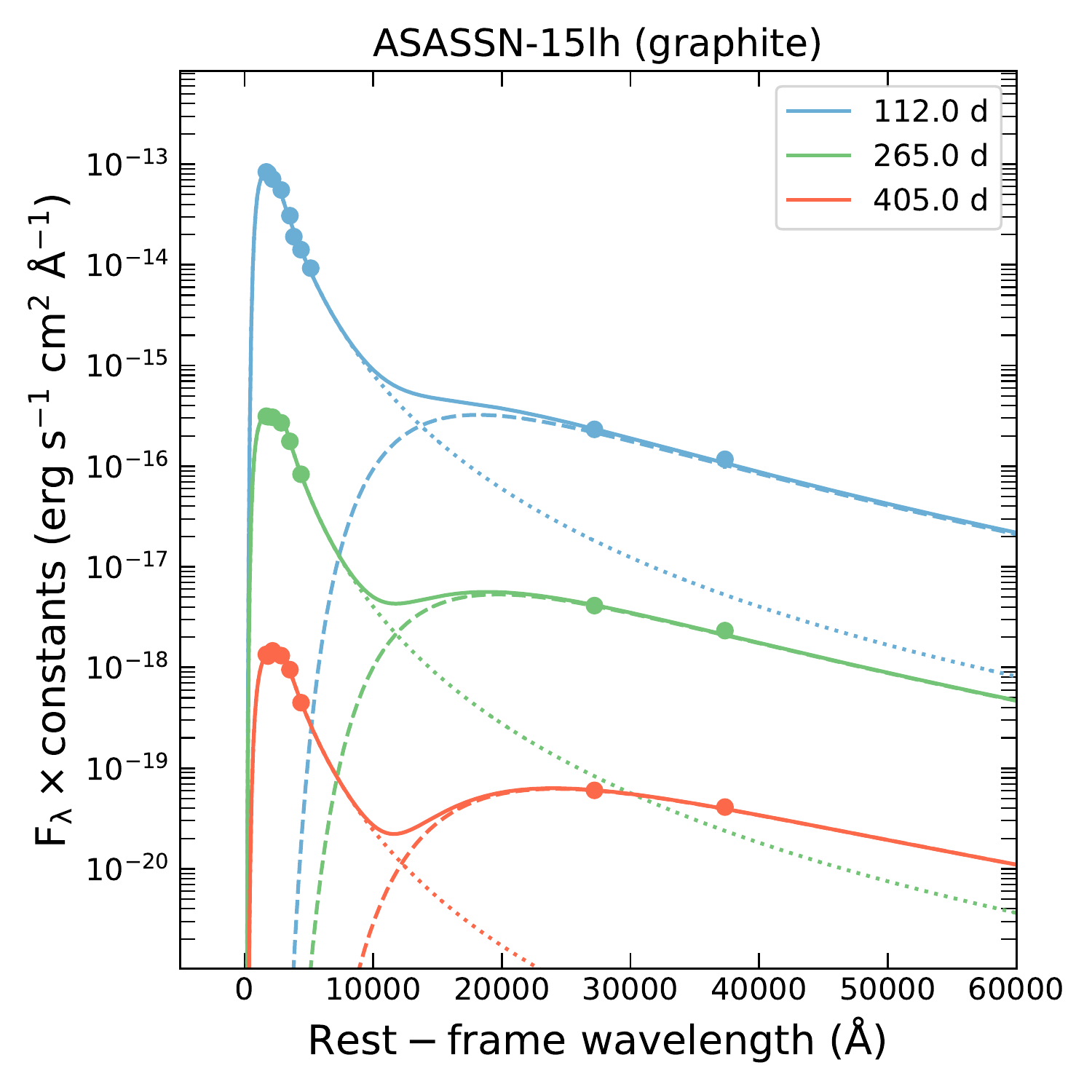}
\includegraphics[width=0.25\textwidth,angle=0]{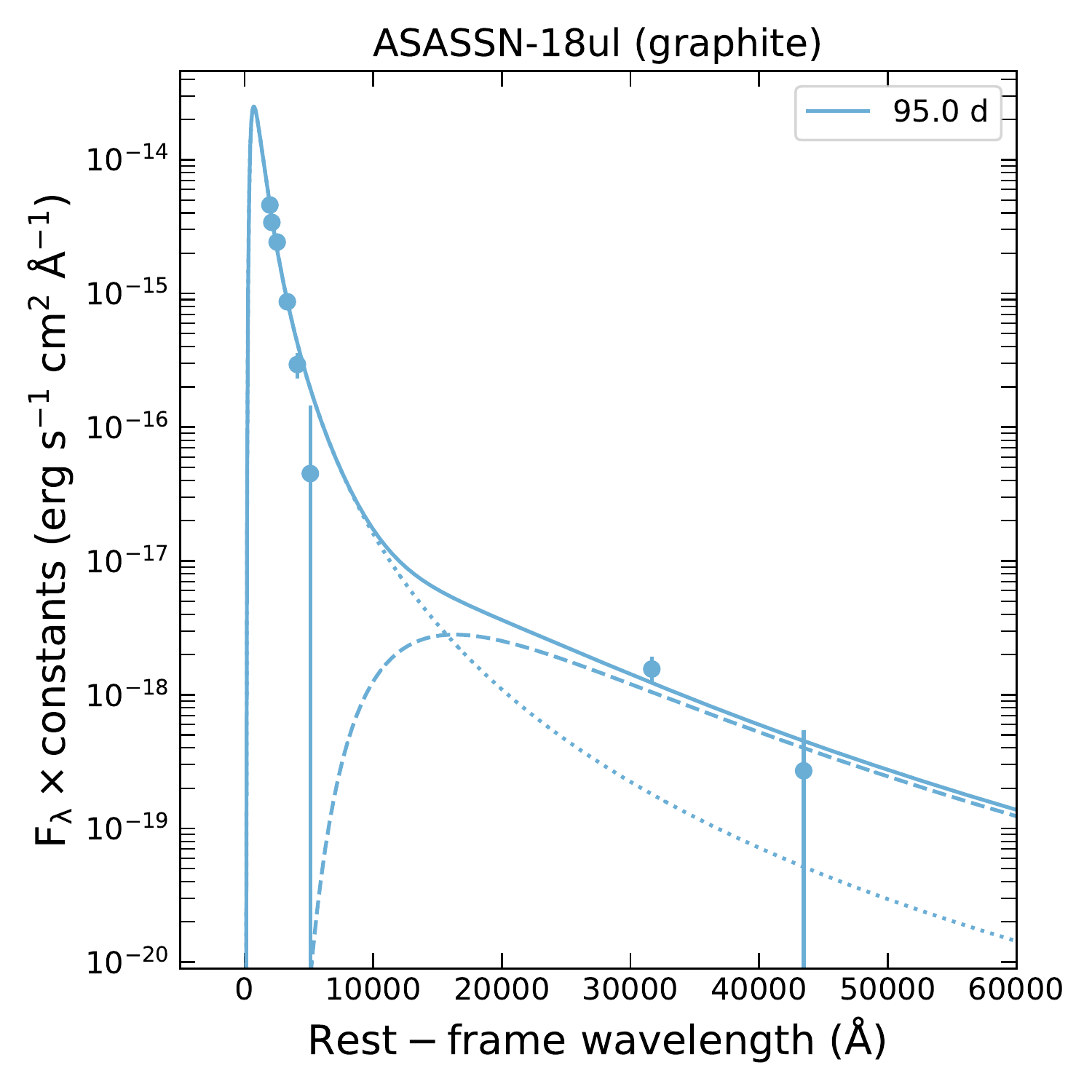}
\includegraphics[width=0.25\textwidth,angle=0]{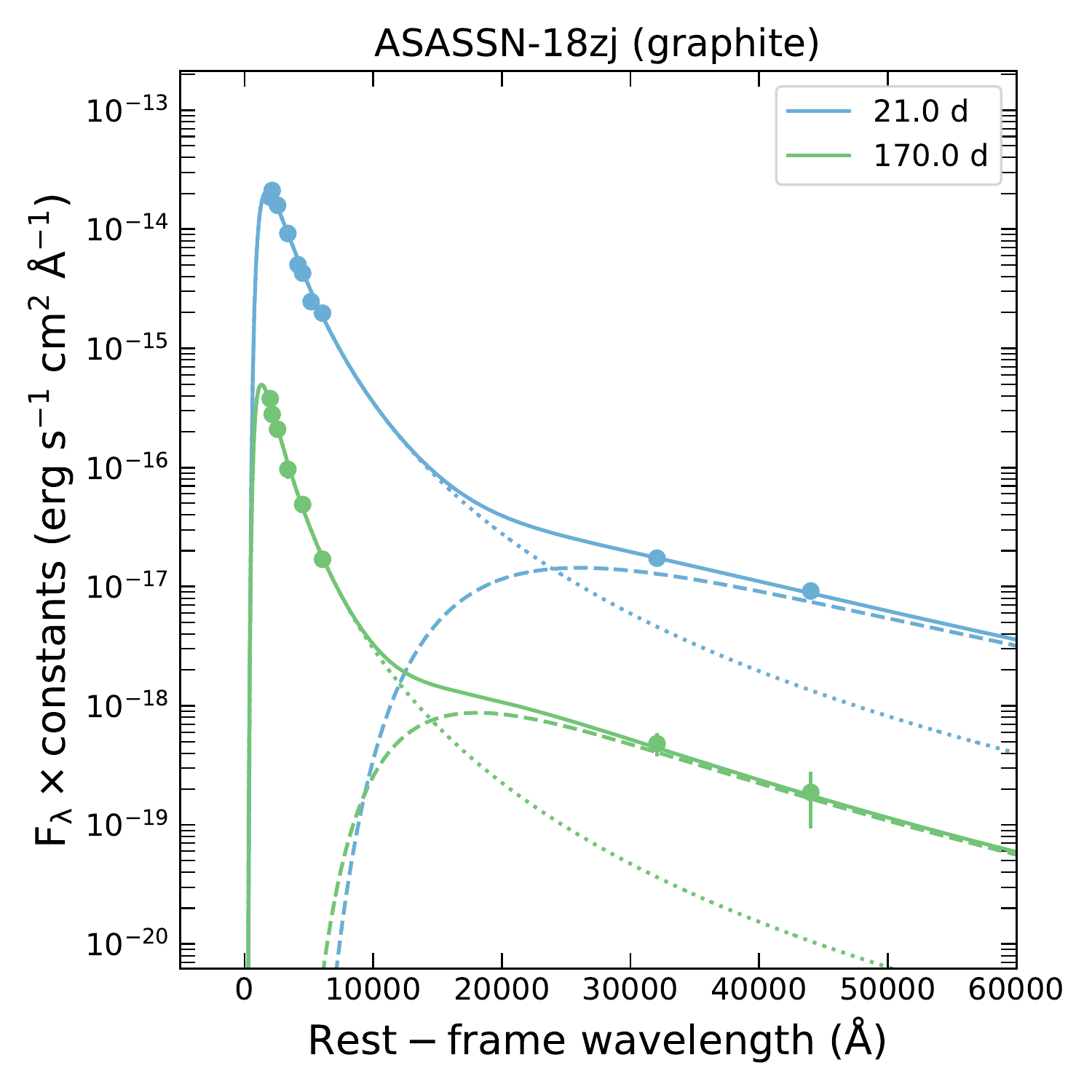}
\includegraphics[width=0.25\textwidth,angle=0]{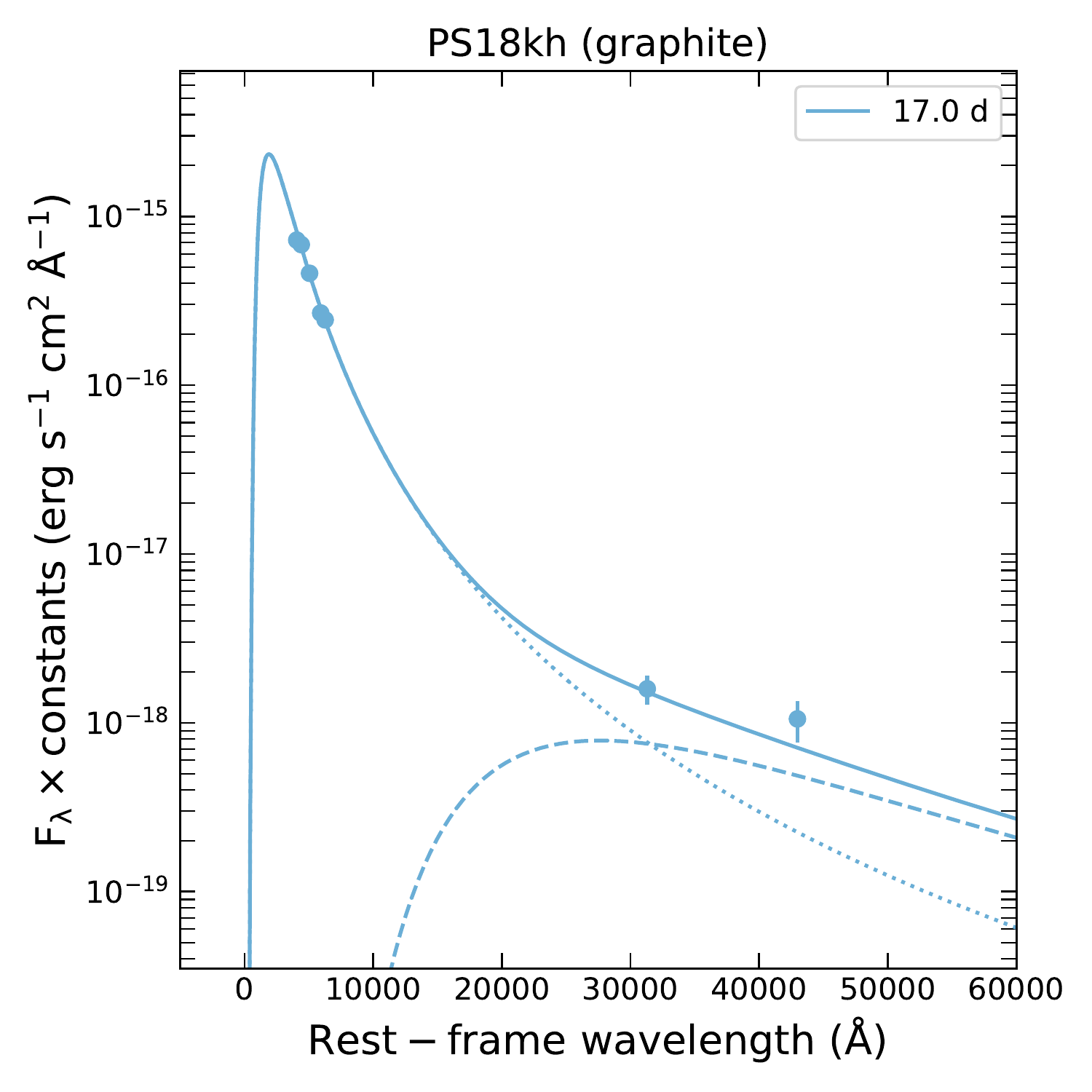}
\includegraphics[width=0.25\textwidth,angle=0]{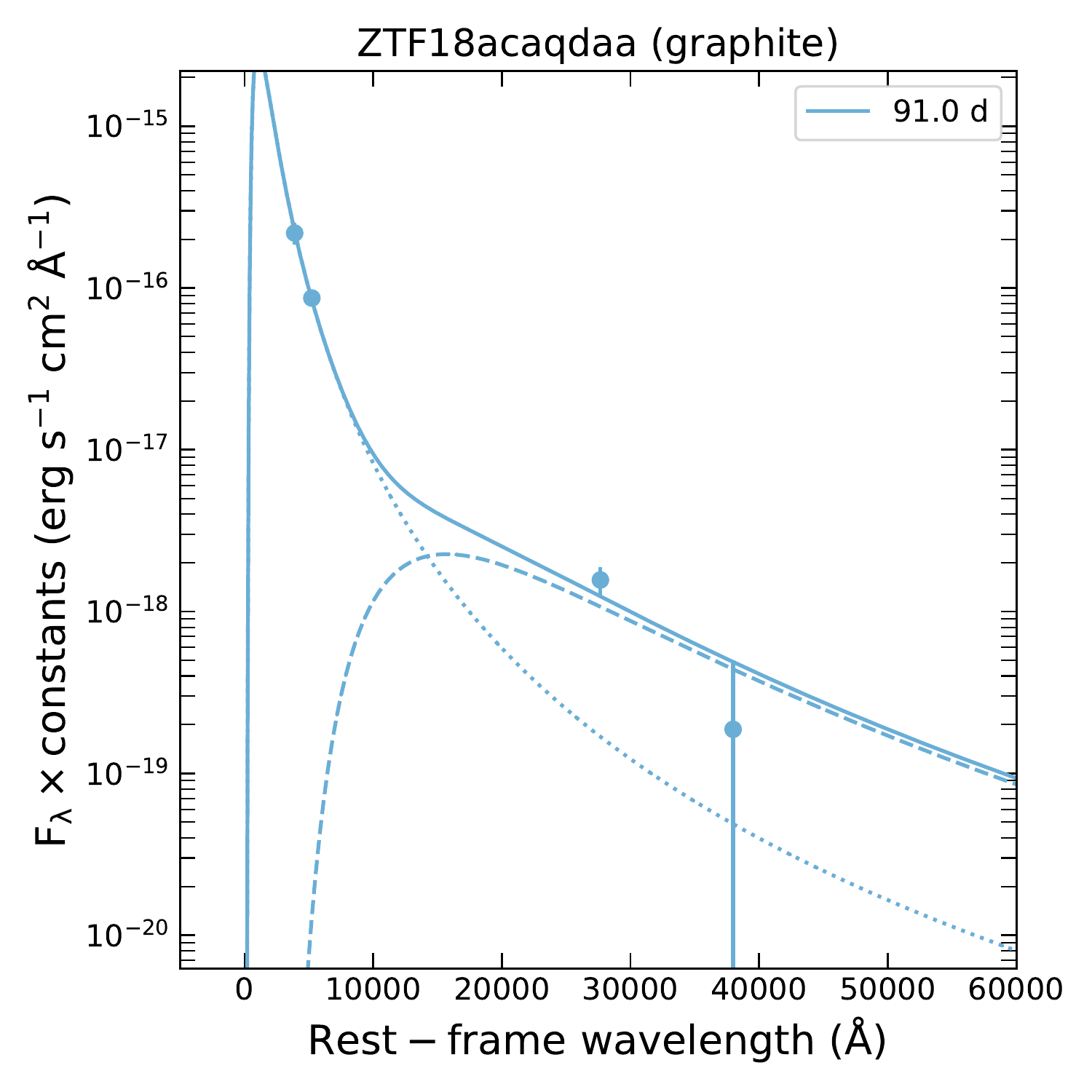}
\includegraphics[width=0.25\textwidth,angle=0]{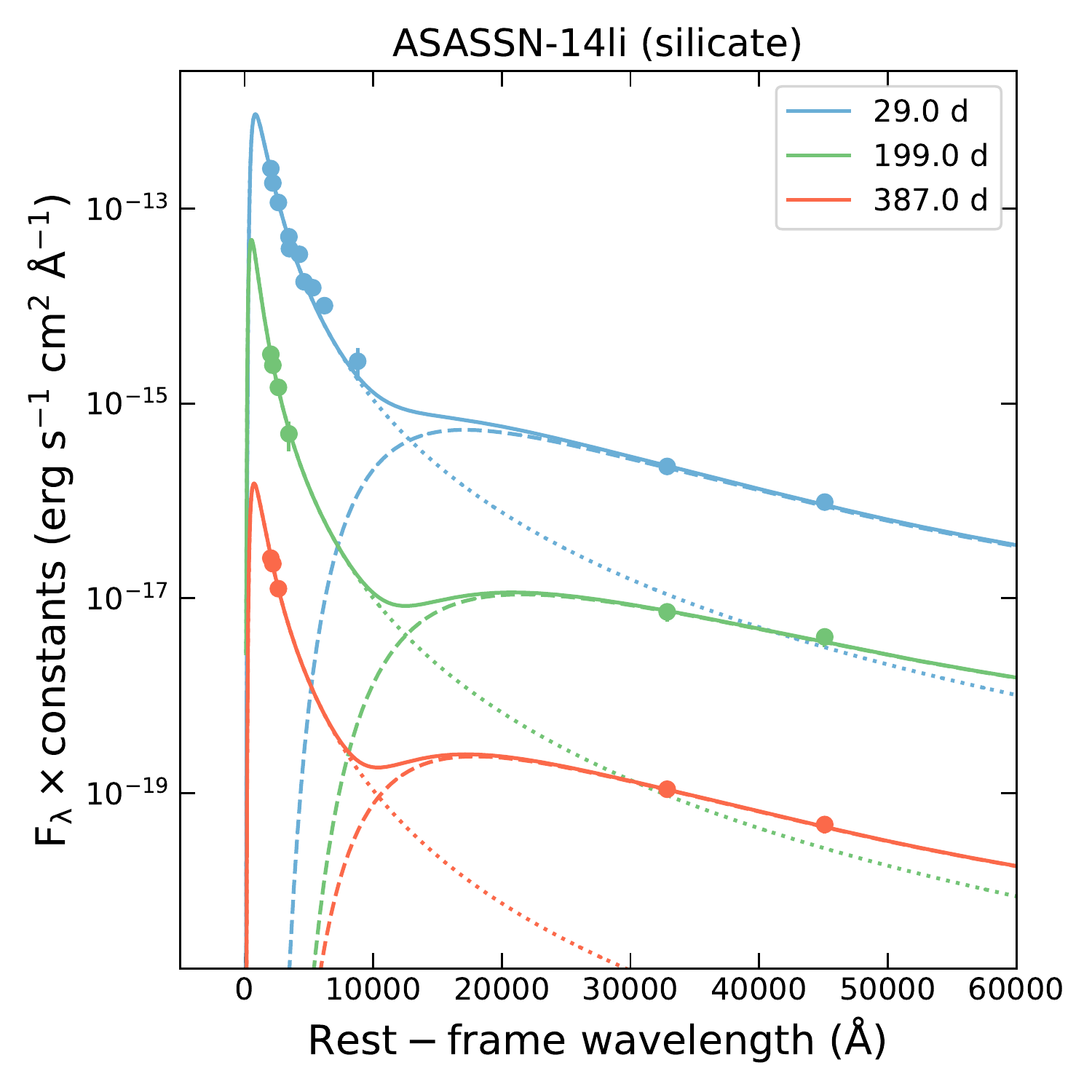}
\includegraphics[width=0.25\textwidth,angle=0]{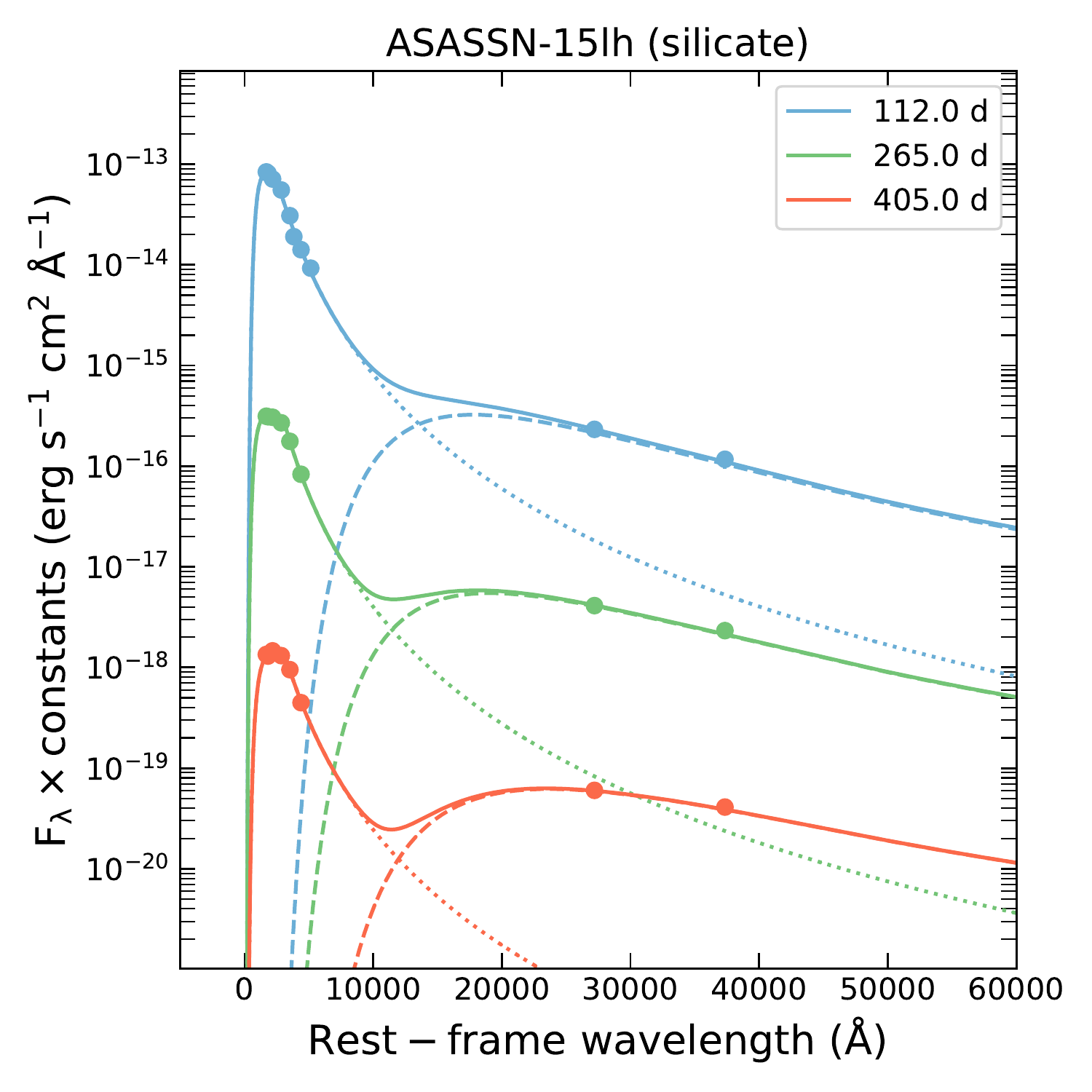}
\includegraphics[width=0.25\textwidth,angle=0]{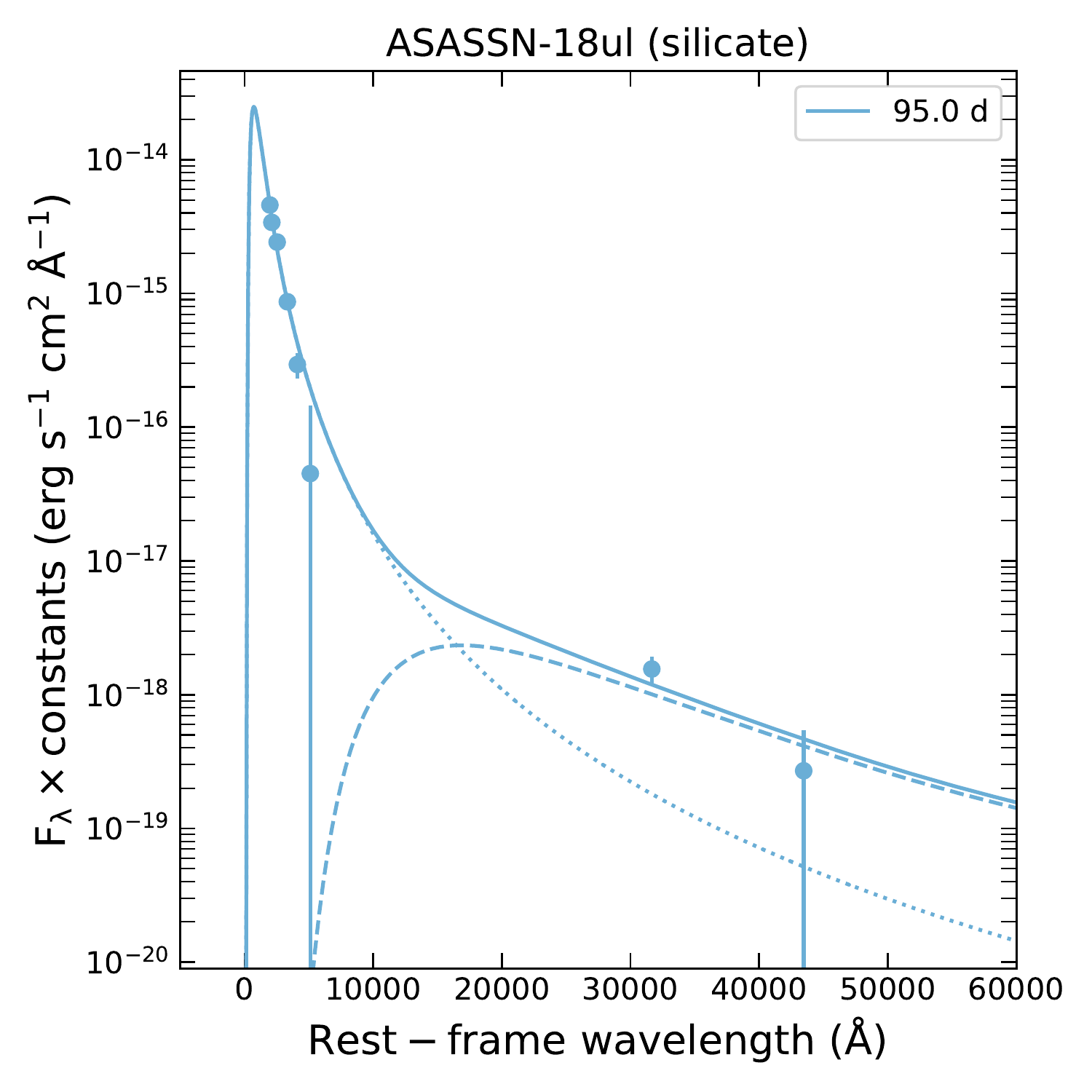}
\includegraphics[width=0.25\textwidth,angle=0]{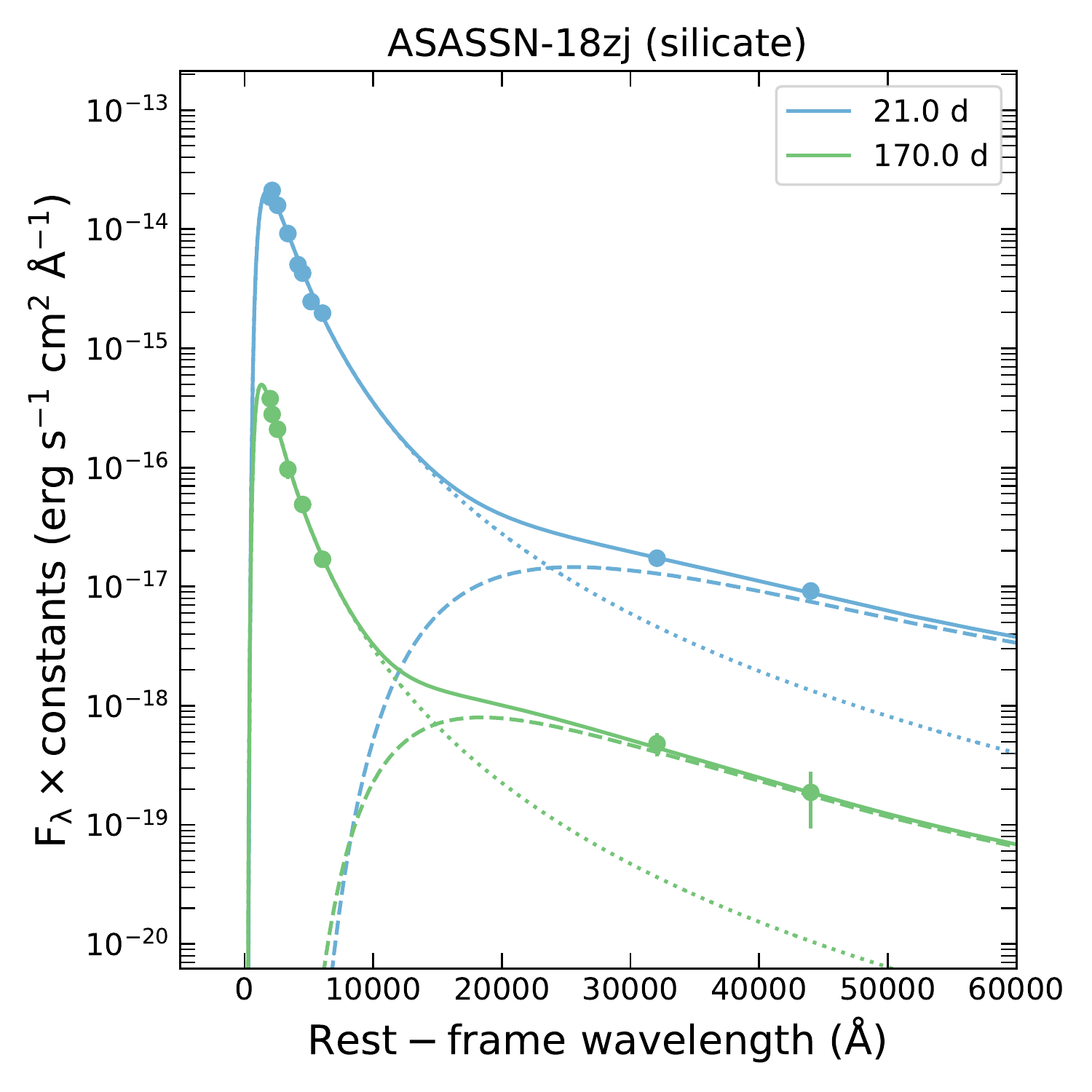}
\includegraphics[width=0.25\textwidth,angle=0]{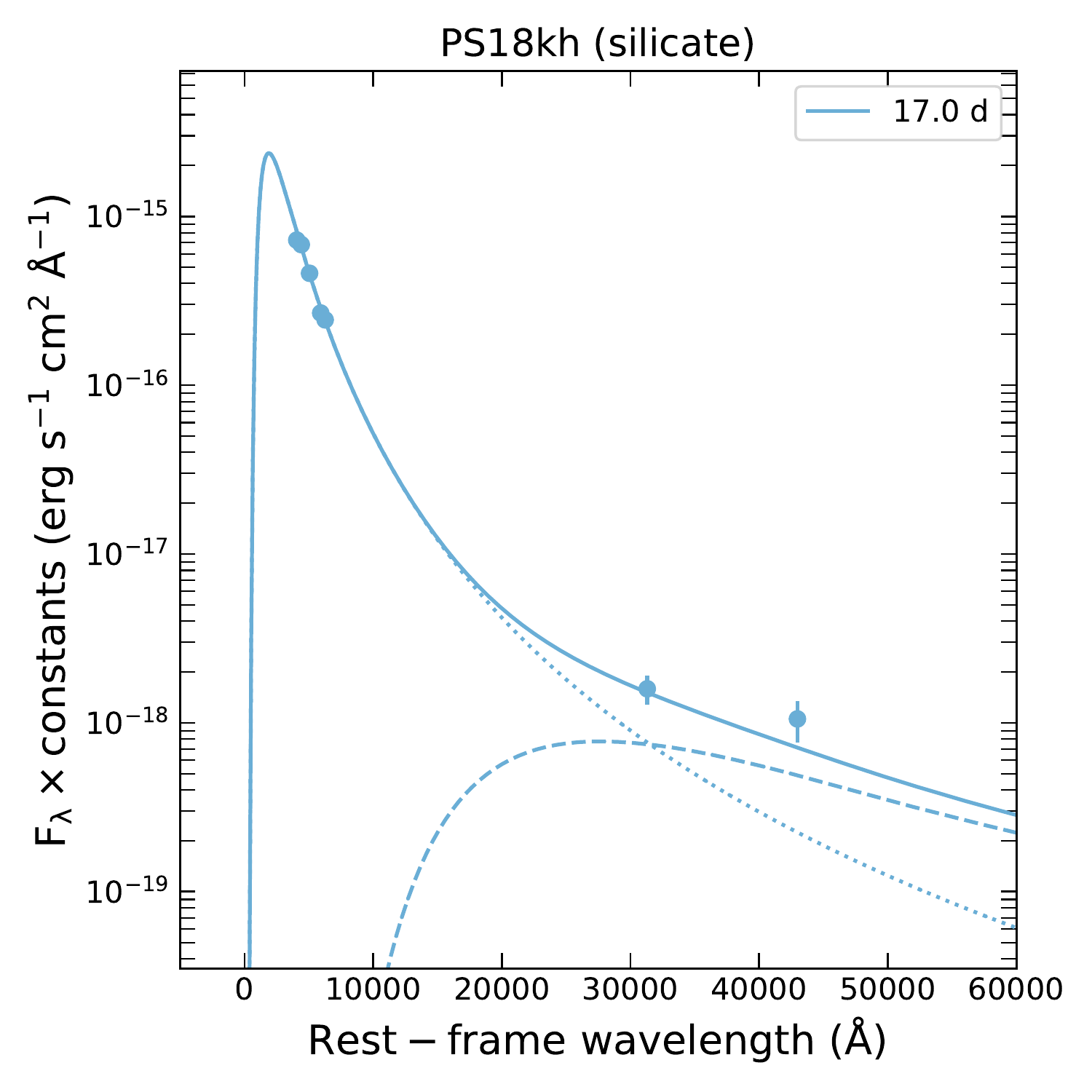}
\includegraphics[width=0.25\textwidth,angle=0]{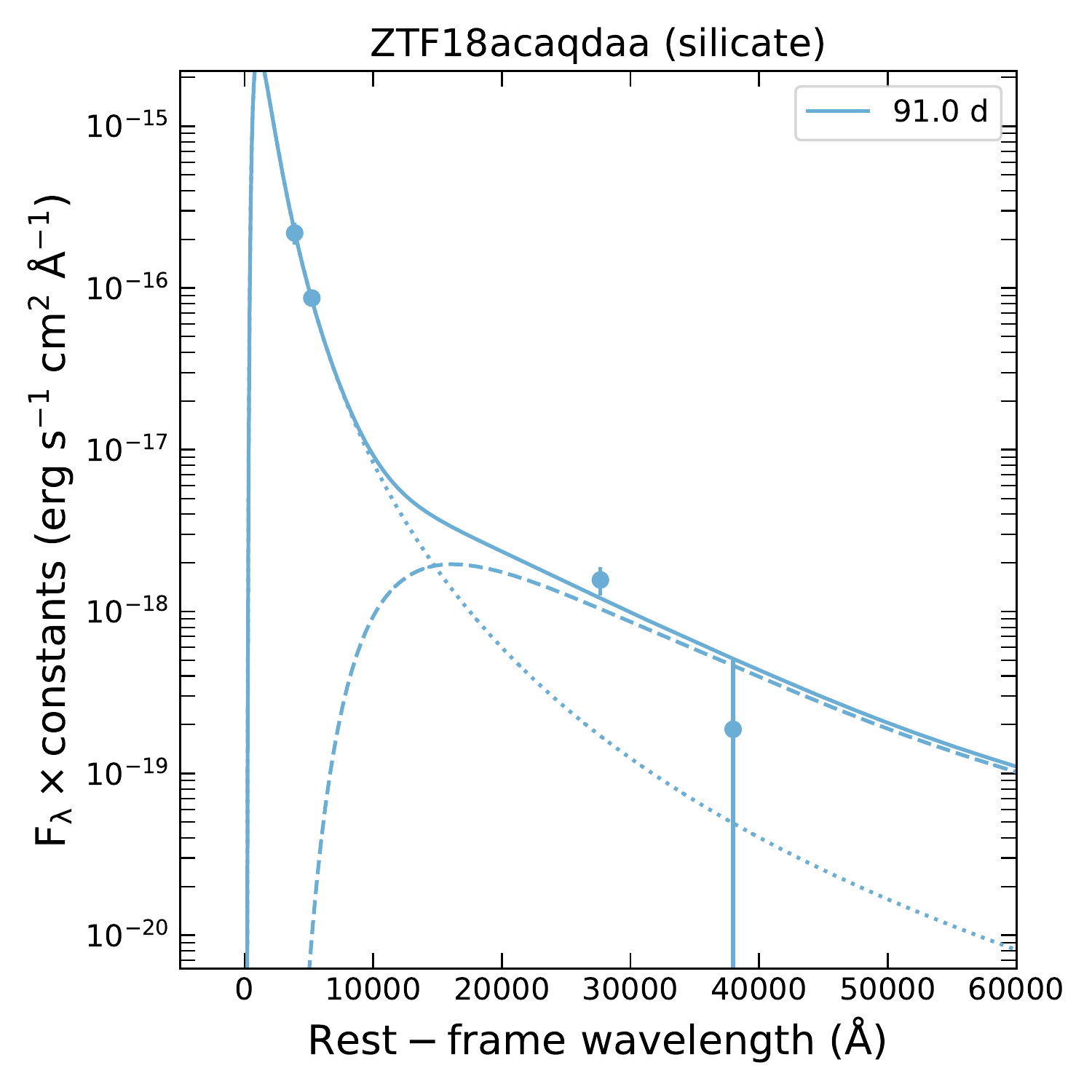}
\end{center}
\caption{The best-fitting SEDs of ASASSN-14li, ASASSN-15lh, ASASSN-18ul, ASASSN-18zj, PS18kh, and ZTF18acaqdaa obtained by using the
blackbody plus graphite (the first two rows) or silicate (the last two rows) model. The dotted, dashed, and solid lines represent
the flux of the photospheres, the dust, and the sum of the two components, respectively.}
\label{fig:SED-dust}
\end{figure}

The best-fitting blackbody plus graphite (or silicate) dust model for the
SEDs and the corresponding best-fitting parameters (masses and temperature) of the model are
shown in Figure \ref{fig:SED-dust} and Table \ref{table:SED-double}, respectively.
We find that the model can explain the SEDs of the six TDEs.
The derived masses of the graphite (silicate) dust of ASASSN-14li, ASASSN-15lh, ASASSN-18ul, ASASSN-18zj,
PS18kh, and ZTF18acaqdaa are
$\sim0.7-1.0\,(1.5-2.2)\times10^{-4}\,M_\odot$,
$\sim0.6-3.1\,(1.4-6.3)\times10^{-2}\,M_\odot$,
$\sim1.0\,(2.8)\times10^{-4}\,M_\odot$,
$\sim0.1-1.6\,(0.3-3.3)\times10^{-3}\,M_\odot$,
$\sim1.0\,(2.0)\times10^{-3}\,M_\odot$, and
$\sim 1.1\,(2.9)\times10^{-3}\,M_\odot$, respectively.
The derived temperatures of the graphite (silicate) dust of the six TDEs
are $\sim1140-1430\,(1210-1520)$\,K, $\sim1030-1380\,(1100-1460)$\,K, $\sim1530\,(1540)$\,K,
$\sim960-1380\,(1020-1420)$\,K, $\sim900\,(950)$\,K, and $\sim1600\,(1610)$\,K, respectively.
The derived dust masses of the silicate are about two to three times those of the graphite
dust. Moreover, the derived mass of the dust of ASASSN-15lh is $(0.6-3.1)\,(1.4-6.3)\times10^{-2}\,M_\odot$,
which is one or two magnitudes larger than those of other five TDEs.

\begin{figure}
\centering
\includegraphics[width=0.40\textwidth,angle=0]{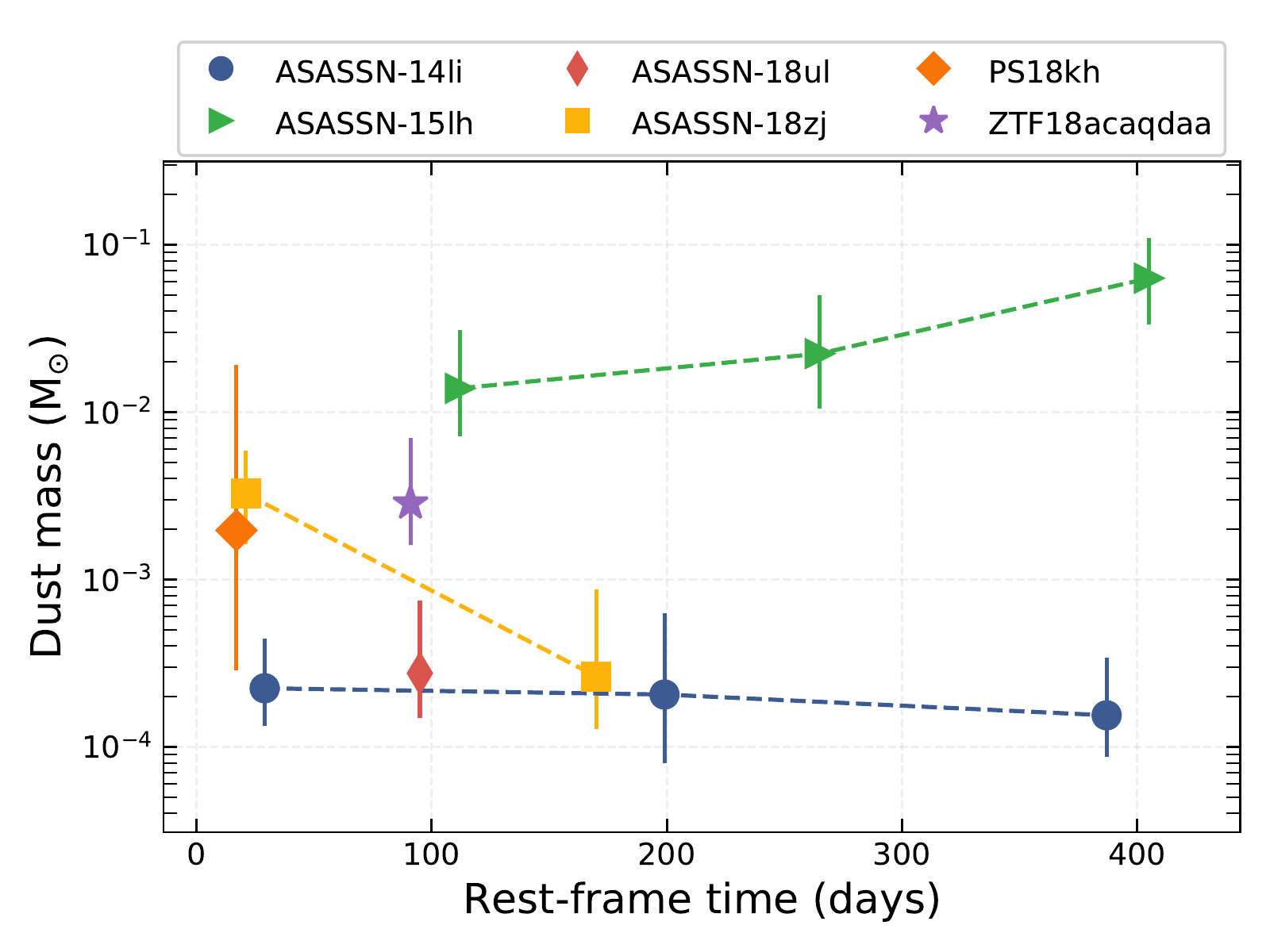}
\includegraphics[width=0.40\textwidth,angle=0]{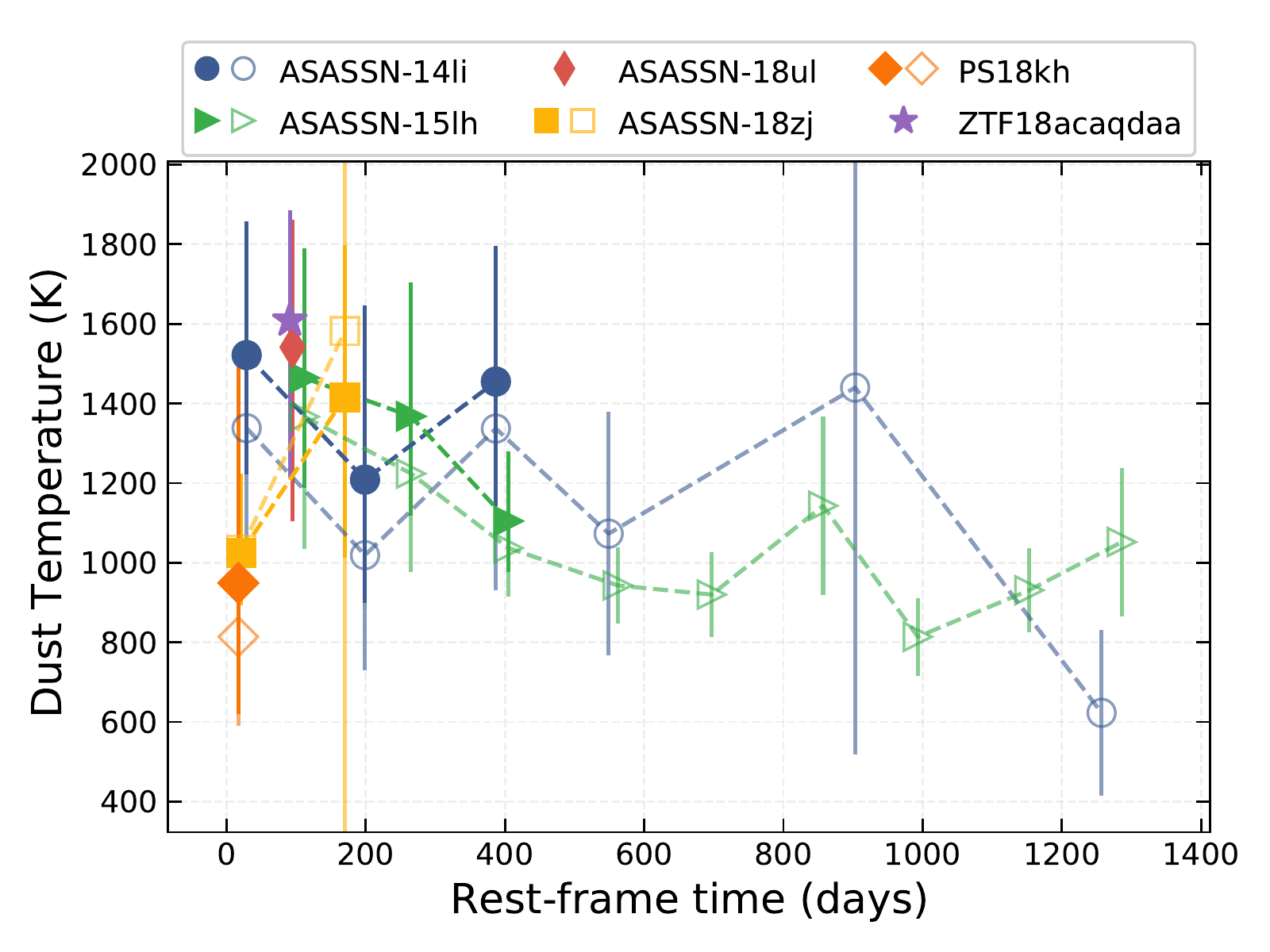}
\caption{The masses (left panel), the temperatures (right panel) of the dust of the six TDEs.
For comparison, the temperatures of the dust derived by \cite{Jiang2021} for four TDEs (ASASSN-14li, ASASSN-15lh, PS18kh, and ASASSN-18zj)
are also plotted by the open symbols in the right panel.}
\label{fig:evolution-M&T}
\end{figure}

We plot the evolution of the masses and the temperatures of the dust in Figure \ref{fig:evolution-M&T}.
For comparison, the temperatures of the dust of four TDEs (ASASSN-14li, ASASSN-15lh, PS18kh, ASASSN-18zj)
derived by \cite{Jiang2021} are also presented in the same Figure \ref{fig:evolution-M&T}.

We find that the derived temperatures of graphite dust of the six TDEs
are reasonable, since they are lower than the vaporization temperature of
graphite $T_{\rm vap,g}$ ($\sim$ 1900 K, e.g., \citealt{Stritzinger2012}).
The silicate dust cannot explain the IR excesses of ASASSN-18ul and ZTF18acaqdaa
since the derived temperatures are higher than
the vaporization temperature of silicate $T_{\rm vap,s}$
($\sim 1100-1300$\,K, \citealt{Mattila2008}; or $\sim$ 1500 K, \citealt{Gall2014});
the IR excesses of ASASSN-14li, ASASSN-15lh and ASASSN-18zj can be narrowly
explained by the silicate dust since the derived temperatures are around the
upper limit of $T_{\rm vap,s}$. IR excess of PS18kh can be due to the silicate
dust emission. Due to the uncertainty of the lower limit of $T_{\rm vap,s}$,
we favor the graphite model, though the silicate model cannot be excluded for
some events.

\section{Discussion}
\label{sec:discussion}

\subsection{The luminosity and the lower limit of the radii of the dust shells}

\begin{figure}
\centering
\includegraphics[width=0.40\textwidth,angle=0]{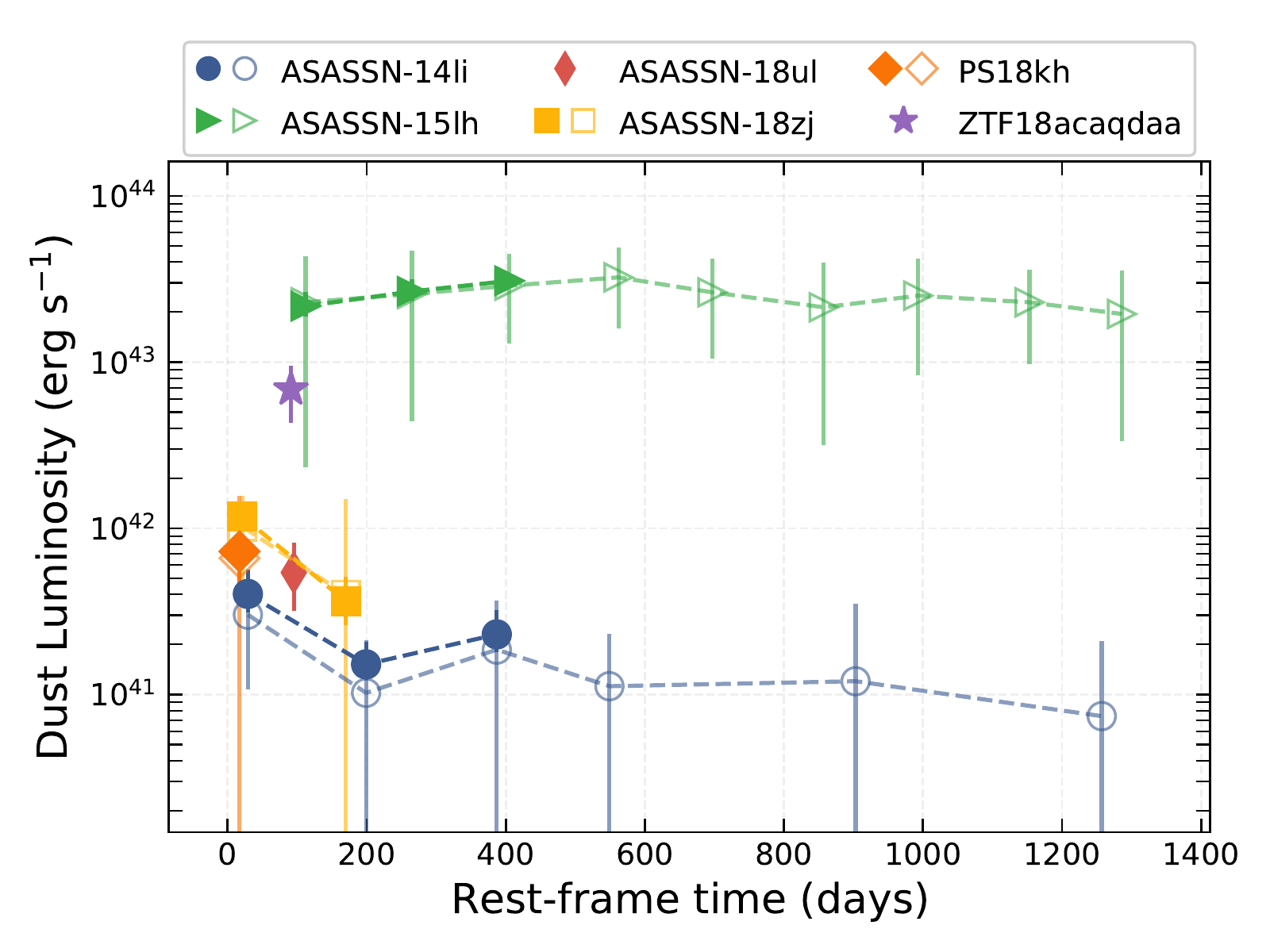}
\includegraphics[width=0.40\textwidth,angle=0]{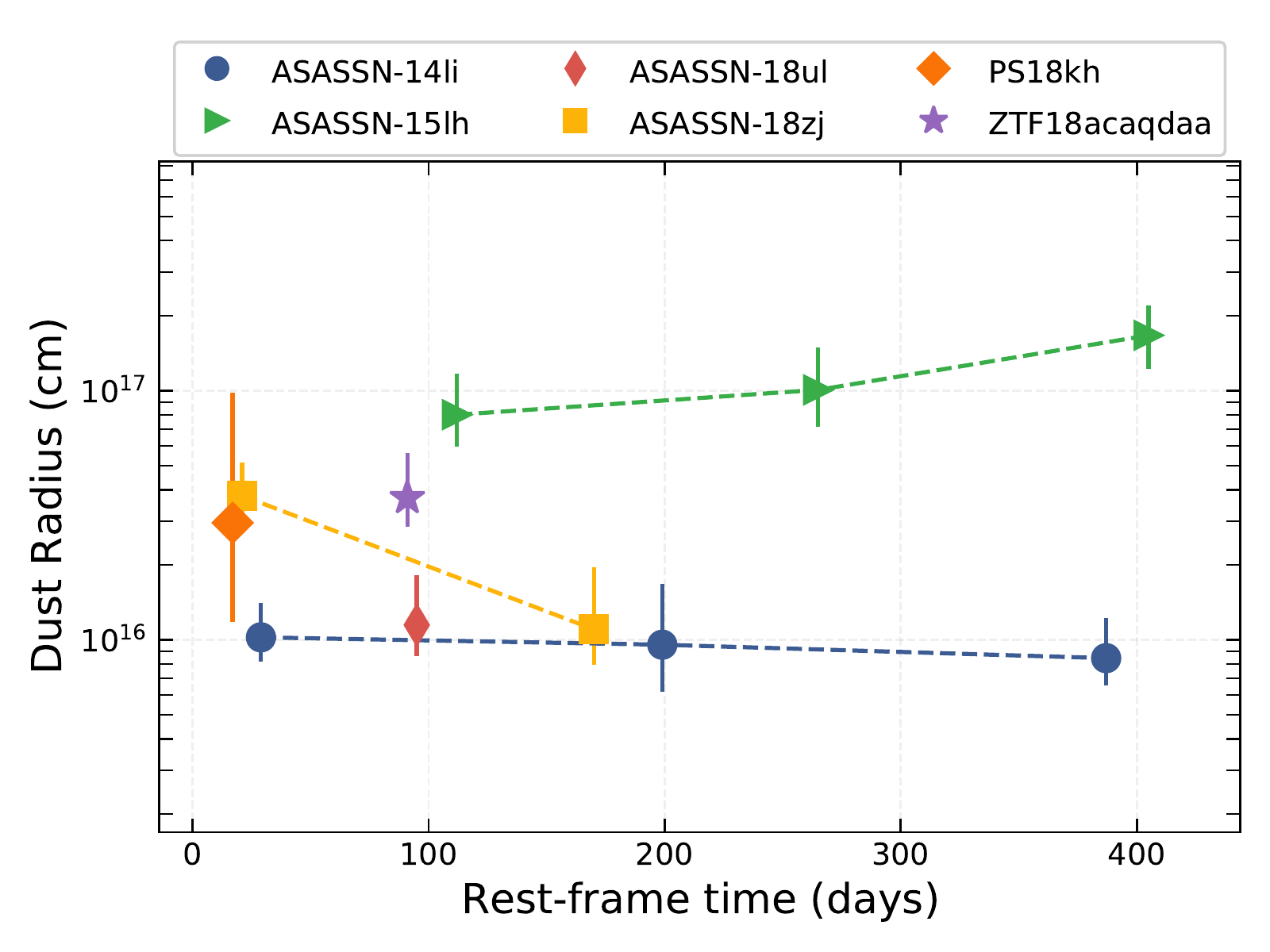}
\caption{The luminosities (left panel) and the radii (right panel) of the dust shells of the six TDEs.
For comparison, the dust luminosities derived by \cite{Jiang2021} for four TDEs (ASASSN-14li, ASASSN-15lh, PS18kh, and ASASSN-18zj)
are also plotted by the open symbols in the left panel.}
\label{fig:evolution-L&R}
\end{figure}

Using the equations in \cite{Fox2011}, we calculate the luminosity
and the blackbody radii ($R_{\rm bb,d}$) of the dust shells which are
the lower limits of the dust shell radii, see Table \ref{table:SED_L}
and Figure \ref{fig:evolution-L&R}.

By comparing $R_{\rm ph}$ and $R_{\rm bb,d}$ of the TDEs at the first epochs when the optical-IR SEDs
are available, we can infer the origin of the dust.
Combining Tables \ref{table:SED_PARAM-single} and \ref{table:SED_L}, we find that $R_{\rm ph}$ ($R_{\rm bb,d}$) of ASASSN-14li
at day 29,  ASASSN-15lh at day 112, ASASSN-18ul at day 95, ASASSN-18zj at day 21, PS18kh
at day 17, ZTF18acaqdaa at day 91 are $\sim1.9\times 10^{14} $ cm ($\sim1.1\times 10^{16}$ cm),
$\sim 2.2\times10^{15}$ cm ($\sim8.4\times10^{16}$ cm), $\sim 3.7\times10^{14}$ cm ($\sim1.2\times10^{16}$ cm),
$\sim 1.4\times10^{15}$ cm ($\sim3.8\times10^{16}$ cm), $\sim1.6\times 10^{15} $ cm ($\sim2.8\times 10^{16} $ cm),
$\sim 4.8\times10^{15}$ cm ($\sim3.8\times10^{16}$ cm), respectively.
The $R_{\rm bb,d}$ values are respectively 58, 38, 32, 27, 18, and 8 times the $R_{\rm ph}$ values for
ASASSN-14li, ASASSN-15lh, ASASSN-18ul, ASASSN-18zj, PS18kh, and ZTF18acaqdaa, indicating that
the dust shells are formed prior the TDE occurred.

\subsection{Comparison to the results in the literature}

\cite{Jiang2021} assume that the dust is silicate and find that the temperature
of ASASSN-14li and ASASSN-15lh is $\sim$1400 K, the temperature of PS18kh and ASASSN-18zj is $\sim$1000 K.
Our derived temperatures of the dust of the four TDEs are consistent with those of \cite{Jiang2021},
see the right panel of Figure \ref{fig:evolution-M&T}.
This supports the conclusion that the four TDEs show evident IR excesses.
Our derived luminosities of the dust of the four TDEs are also consistent with those of \cite{Jiang2021},
see the left panel of Figure \ref{fig:evolution-L&R}.
Additionally, we obtain the temperatures and luminosities of the dust of ASASSN-18ul and ZTF18acaqdaa,
while they are absent in the literature.

The consistency between our results and \cite{Jiang2021}'s results might be
due to the fact that the two assumptions adopted by \cite{Jiang2021} are valid,
or the fact that the IR flux from the hot photospheres of the TDEs are (significantly) lower than
that from the cool dust and the results are insensitive to the bolometric light-curve and temperature
evolution patterns assumed.

\section{Conclusions}
\label{sec:Con}

In this paper, we collect the (UV-)optical and IR data of six TDEs
(ASASSN-14li, ASASSN-15lh, ASASSN-18ul, ASASSN-18zj, PS18kh, and ZTF18acaqdaa)
from the literature, constructing and fitting their optical-IR SEDs at different epochs.
The blackbody model fits show that the SEDs of the all six TDEs
show IR excesses, while the blackbody plus dust emission model can
account for the SEDs showing IR excesses.

The fits using the blackbody plus dust emission model show that the IR excesses of ASASSN-14li,
ASASSN-15lh, ASASSN-18ul, ASASSN-18zj, PS18kh, and ZTF18acaqdaa are produced by the graphite (silicate)
dust of the masses of $\sim0.7-1.0\,(1.5-2.2)\times10^{-4}\,M_\odot$,
$\sim0.6-3.1\,(1.4-6.3)\times10^{-2}\,M_\odot$,
$\sim1.0\,(2.8)\times10^{-4}\,M_\odot$,
$\sim0.1-1.6\,(0.3-3.3)\times10^{-3}\,M_\odot$,
$\sim1.0\,(2.0)\times10^{-3}\,M_\odot$, and
$\sim 1.1\,(2.9)\times10^{-3}\,M_\odot$, respectively.
The derived temperatures of the graphite (silicate) dust of them
are $\sim1140-1430\,(1210-1520)$\,K, $\sim1030-1380\,(1100-1460)$\,K, $\sim1530\,(1540)$\,K,
$\sim960-1380\,(1020-1420)$\,K, $\sim900\,(950)$\,K, and $\sim1600\,(1610)$\,K, respectively.

By comparing the derived temperatures to the vaporization temperature
of graphite ($\sim 1900$\,K) and silicate ($\sim 1100-1500$\,K),
we suggest that the IR excesses of PS18kh can be explained by both the
graphite and silicate dust, the rest five TDEs favor the
graphite dust while the silicate dust model cannot be excluded.

Moreover, we calculate the dust luminosities ($L_{\rm ph}$) and the blackbody radii ($R_{\rm bb,d}$)
which are the lower limits of the dust shells, finding that
$R_{\rm bb,d}$ of the six TDEs are tens or hundreds of times $R_{\rm ph}$ at the epochs when their first SEDs
are available. This indicates that the IR excesses of the TDEs are produced by the pre-existing dust.

The advantage of our method is that it does not depend on any assumptions of evolution patterns of
bolometric luminosities and temperatures. On the other hand, however, its disadvantage is that they are fewer
epochs when (UV-)optical and IR photometry are available. We expected that (UV-)optical and IR observations for
TDEs in the future can provide more (UV-)optical-IR SEDs for individual events that can be used to constrain
the properties of the dust.

\acknowledgments
This work is supported by National Natural Science Foundation of China
(grant 11963001).

\clearpage

\setlength{\tabcolsep}{2pt}
\begin{table*}
\begin{center}
\footnotesize\caption{The information of the six TDEs in the sample.}
\label{table:details}
\begin{tabular}{cccccccccccc}
\hline
\hline
Name & R.A.  & Decl. & Redshift &  Discovery date & Obs. Filters & References$^a$ \\
  & (J2000)  &  (J2000)   &   &    &   &   \\
\hline
\hline
ASASSN-14li   & \ra{12}{48}{15}{23}& \dec{+17}{46}{26}{4}  & 0.0206 & Nov 22 2014 & $UVW2$, $UVM2$, $UVW1$, $r$, $B$, $V$, $g$, $U$, $i$, $W_1$, $W_2$  & 1, 2 \\
ASASSN-15lh  & \ra{22}{02}{15}{39}& \dec{-61}{39}{34}{6}  & 0.2326  & Jun 14 2015 & $UVW2$, $UVM2$, $UVW1$, $g$, $i$, $r$, $B$, $V$, $U$, $W_1$, $W_2$  & 1, 3, 4 \\
ASASSN18ul   & \ra{22}{50}{16}{064}& \dec{-44}{51}{53}{26}  & 0.059 & Sep 08 2018 & $UVW2$, $UVM2$, $UVW1$, $U$, $B$, $V$, $W_1$, $W_2$ & 1, 5  \\
ASASSN18zj    & \ra{10}{06}{50}{871}& \dec{+01}{41}{34}{08}  & 0.04573 & Dec 14 2018  & $UVW2$, $UVM2$, $UVW1$, $U$, $B$, $g$, $r$, $i$, $W_1$, $W_2$  & 1, 6\\
PS18kh    & \ra{07}{56}{54}{54}& \dec{+34}{15}{43}{6}  & 0.0710 & Feb 07 2018  & $g$, $r$, $o$, $B$, $V$, $W_1$, $W_2$  & 1, 7 \\
ZTF18acaqdaa    & \ra{17}{28}{03}{93}& \dec{+30}{41}{31}{4}  & 0.2120 & Nov 11 2018 & $g$, $r$, $W_1$, $W_2$  & 1, 8 \\
\hline
\hline
\end{tabular}
\end{center}
$^a$ References. (1) \cite{Jiang2021}; (2)\cite{Holoien2016}; (3) \cite{Godoy2017}; (4) \cite{Leloudas2016};
(5) \cite{Wevers2019}; (6) \cite{Gomez2020}; (7) \cite{Holoien2019}; (8) {\it MARS} (https://mars.lco.global). \\
\end{table*}

\clearpage

 \begin{table*}
 	\renewcommand{\arraystretch}{1.5}
 	\caption{\label{table:SED_PARAM-single}The best-fitting parameters of the blackbody model for the SEDs of the six TDEs, including the temperature of the photosphere $T_{\rm ph}$, the radius of the photosphere $R_{\rm ph}$}.
 	\centering
 	\begin{tabular}{c c c c c c c c c c c c c}
 		\hline\hline\noalign{\smallskip}
 		\colhead{Phase\footnote{All the phases are relative to the peak date which are respectively MJD 56989 (ASASSN-14li), MJD 57247 (ASASSN-15lh), MJD 58317 (ASASSN-18ul),
 MJD 58180 (PS18kh), MJD 58427 (ASASSN-18zj), and MJD 58442 (ZTF18acaqdaa) (see \citealt{Jiang2021}). All epochs are transformed to the rest-frame ones.}} & \colhead{$T_{\rm ph}$}&\colhead{$R_{\rm ph}$} & \colhead{$\chi^{\rm 2}$/dof} \\
 		(days) & (10$^{3}$\,K) & (10$^{15}$ cm)    & \\\hline
 		\multicolumn{4}{c}{\textbf{ASASSN-14li}}\\\hline
 		29.0 d  &  $34.00^{+1.34}_{-1.21}$  &  $0.19^{+0.01}_{-0.01}$  &  21.95 \\\hline
 		199.0 d  &  $51.22^{+25.09}_{-14.83}$  &  $0.051^{+0.020}_{-0.014}$  &  10.84 \\\hline
 		387.0 d  &  $33.13^{+22.21}_{-9.40}$  &  $0.072^{+0.040}_{-0.029}$  &  26.43 \\\hline
 		\multicolumn{4}{c}{\textbf{ASASSN-15lh}}\\\hline
 		112.0 d  &  $20.06^{+3.15}_{-1.46}$  &  $2.23^{+0.21}_{-0.31}$  &  15.88 \\\hline
 		265.0 d  &  $27.97^{+23.92}_{-8.70}$  &  $1.02^{+0.47}_{-0.40}$  &  25.01 \\\hline
 		405.0 d  &  $19.26^{+23.91}_{-4.58}$  &  $0.96^{+0.41}_{-0.50}$  &  44.99 \\\hline
 		
 		\multicolumn{4}{c}{\textbf{ASASSN-18ul}}\\\hline
 		95.0 d  &  $39.41^{+5.10}_{-3.87}$  &  $0.37^{+0.04}_{-0.04}$  &  6.06 \\\hline
 		
 		\multicolumn{4}{c}{\textbf{ASASSN-18zj}}\\\hline
 		
 		21.0 d  &  $16.20^{+0.19}_{-0.19}$  &  $1.40^{+0.03}_{-0.03}$  &  30.97 \\\hline
 		170.0 d  &  $21.91^{+1.16}_{-1.08}$  &  $0.30^{+0.02}_{-0.02}$  &  4.09 \\\hline
 		
 		\multicolumn{4}{c}{\textbf{PS18kh}}\\\hline
 		17.0 d  &  $14.15^{+1.53}_{-1.25}$  &  $1.62^{+0.18}_{-0.17}$  &  3.54 \\\hline
 		
 		\multicolumn{4}{c}{\textbf{ZTF18acaqdaa}}\\\hline
 		91.0 d  &  $8.99^{+6.95}_{-1.93}$  &  $4.80^{+2.68}_{-2.61}$  &  7.65 \\\hline

 	\end{tabular}
 \end{table*}

 \clearpage

 \begin{table*}
 	\centering
 	\footnotesize
 	\setlength{\tabcolsep}{1.8pt}
 	\renewcommand{\arraystretch}{1.5}
 	\caption{\label{table:SED-double}The best-fitting parameters (the temperature of the photosphere $T_{\rm ph}$, the radius of the photosphere $R_{\rm ph}$, the temperature of the dust shell $T_{\rm d}$, and the mass of the dust shell $M_{\rm d}$) of the blackbody plus graphite (silicate) model for the SEDs of the six TDEs.}
 	\centering
 	\begin{tabular}{c c c c c c c c c c c c c}
 		\hline\hline\noalign{\smallskip}
 		&\multicolumn{5}{c}{\textbf{Blackbody plus Graphite}} & \multicolumn{5}{c}{\textbf{Blackbody plus Silicate}}\\\noalign{\smallskip}
 		\cmidrule(lr){2-6} \cmidrule(lr){7-11}
 		\colhead{Phase} & \colhead{$T_{\rm ph}$}&\colhead{$R_{\rm ph}$}&\colhead{$T_{\rm d}$}&\colhead{$M_{\rm d}$}&\colhead{$\chi^{\rm 2}$/dof}&\colhead{$T_{\rm ph}$}&\colhead{$R_{\rm ph}$}&\colhead{$T_{\rm d}$} & \colhead{$M_{\rm d}$}  & \colhead{$\chi^{\rm 2}$/dof} \\
 		(days) & (10$^{3}$\,K) & (10$^{15}$ cm) &  (10$^{3}$\,K) & (10$\rm ^{-3}\ M_{\odot}$) &    & (10$^{3}$\,K) & (10$^{15}$ cm) & (10$^{3}$\,K) & (10$\rm^{-3}\ M_{\odot}$) & \\\hline
 		\multicolumn{11}{c}{\textbf{ASASSN-14li}}\\\hline
 		29.0 d  &  $34.38^{+1.37}_{-1.26}$  &  $0.19^{+0.01}_{-0.01}$  &  $1.43^{+0.39}_{-0.28}$  &  $0.10^{+0.11}_{-0.05}$  &  6.29  &  $34.37^{+1.37}_{-1.23}$  &  $0.19^{+0.01}_{-0.01}$  &  $1.52^{+0.34}_{-0.30}$  &  $0.22^{+0.22}_{-0.09}$  &  6.26 \\\hline
 		199.0 d  &  $54.10^{+24.24}_{-16.10}$  &  $0.049^{+0.019}_{-0.013}$  &  $1.14^{+0.44}_{-0.28}$  &  $0.095^{+0.201}_{-0.062}$  &  0.40  &  $53.92^{+25.35}_{-15.92}$  &  $0.049^{+0.019}_{-0.013}$  &  $1.21^{+0.44}_{-0.31}$  &  $0.21^{+0.42}_{-0.13}$  &  0.38 \\\hline
 		387.0 d  &  $39.57^{+26.02}_{-12.03}$  &  $0.059^{+0.031}_{-0.022}$  &  $1.38^{+0.37}_{-0.30}$  &  $0.068^{+0.088}_{-0.033}$  &  0.85  &  $39.24^{+25.95}_{-11.89}$  &  $0.059^{+0.032}_{-0.022}$  &  $1.45^{+0.34}_{-0.31}$  &  $0.15^{+0.19}_{-0.07}$  &  0.77 \\\hline
 		
 		\multicolumn{11}{c}{\textbf{ASASSN-15lh}}\\\hline
 		112.0 d  &  $24.33^{+10.87}_{-4.16}$  &  $1.83^{+0.38}_{-0.50}$  &  $1.38^{+0.35}_{-0.27}$  &  $6.08^{+8.45}_{-3.30}$  &  1.24  &  $24.31^{+10.85}_{-4.15}$  &  $1.83^{+0.38}_{-0.50}$  &  $1.46^{+0.33}_{-0.28}$  &  $13.87^{+17.07}_{-6.68}$  &  1.23 \\\hline
 		265.0 d  &  $38.09^{+29.45}_{-13.88}$  &  $0.78^{+0.38}_{-0.26}$  &  $1.27^{+0.34}_{-0.22}$  &  $10.55^{+13.51}_{-6.15}$  &  0.18  &  $38.27^{+29.61}_{-13.96}$  &  $0.78^{+0.38}_{-0.26}$  &  $1.37^{+0.34}_{-0.25}$  &  $22.35^{+27.77}_{-11.85}$  &  0.16 \\\hline
 		405.0 d  &  $29.40^{+34.00}_{-12.25}$  &  $0.62^{+0.47}_{-0.27}$  &  $1.03^{+0.15}_{-0.11}$  &  $31.14^{+22.72}_{-14.17}$  &  0.19  &  $28.82^{+33.52}_{-11.94}$  &  $0.64^{+0.48}_{-0.28}$  &  $1.10^{+0.18}_{-0.13}$  &  $63.10^{+46.81}_{-29.62}$  &  0.30 \\\hline

 		\multicolumn{11}{c}{\textbf{ASASSN-18ul}}\\\hline
 		95.0 d  &  $40.28^{+5.55}_{-4.04}$  &  $0.36^{+0.04}_{-0.04}$  &  $1.53^{+0.32}_{-0.43}$  &  $0.10^{+0.19}_{-0.05}$  &  5.61  &  $40.21^{+5.44}_{-4.05}$  &  $0.36^{+0.04}_{-0.04}$  &  $1.54^{+0.32}_{-0.44}$  &  $0.28^{+0.47}_{-0.13}$  &  5.67 \\\hline
 		
 		\multicolumn{11}{c}{\textbf{ASASSN-18zj}}\\\hline
 		
 		21.0 d  &  $16.55^{+0.21}_{-0.20}$  &  $1.34^{+0.03}_{-0.03}$  &  $0.96^{+0.17}_{-0.11}$  &  $1.59^{+1.26}_{-0.78}$  &  10.59  &  $16.55^{+0.21}_{-0.20}$  &  $1.34^{+0.03}_{-0.03}$  &  $1.02^{+0.20}_{-0.13}$  &  $3.25^{+2.66}_{-1.63}$  &  10.59 \\\hline
 		170.0 d  &  $22.02^{+1.18}_{-1.08}$  &  $0.30^{+0.02}_{-0.02}$  &  $1.38^{+0.40}_{-0.41}$  &  $0.10^{+0.27}_{-0.05}$  &  1.04  &  $22.02^{+1.18}_{-1.10}$  &  $0.30^{+0.02}_{-0.02}$  &  $1.42^{+0.38}_{-0.40}$  &  $0.26^{+0.60}_{-0.14}$  &  1.04 \\\hline
 		
 		\multicolumn{11}{c}{\textbf{PS18kh}}\\\hline
 		17.0 d  &  $15.31^{+1.85}_{-1.49}$  &  $1.49^{+0.18}_{-0.16}$  &  $0.90^{+0.61}_{-0.31}$  &  $0.92^{+8.33}_{-0.81}$  &  1.55  &  $15.38^{+1.86}_{-1.50}$  &  $1.48^{+0.17}_{-0.16}$  &  $0.95^{+0.59}_{-0.33}$  &  $1.96^{+17.13}_{-1.68}$  &  1.55 \\\hline
 		
 		\multicolumn{11}{c}{\textbf{ZTF18acaqdaa}}\\\hline
 		91.0 d  &  $27.48^{+32.60}_{-12.65}$  &  $1.29^{+1.08}_{-0.56}$  &  $1.60^{+0.29}_{-0.40}$  &  $1.06^{+1.65}_{-0.46}$  &  --  &  $26.89^{+32.20}_{-12.39}$  &  $1.31^{+1.12}_{-0.58}$  &  $1.61^{+0.28}_{-0.39}$  &  $2.85^{+4.14}_{-1.24}$  &  -- \\\hline

 	\end{tabular}
 \end{table*}


 \clearpage

 \begin{table*}
 	\renewcommand{\arraystretch}{1.5}
 	\caption{\label{table:SED_L}The derived parameters (the photosphere luminosities ($L_{\rm ph}$), the dust luminosities ($L_{\rm d}$),
 and the blackbody radii of the dust shells ($R_{\rm bb,d}$) of the TDEs of the blackbody plus graphite (silicate) model.}
 	\centering
 	\begin{tabular}{c c c c c c c}
 		\hline\hline\noalign{\smallskip}
 		\colhead{Phase} & \colhead{$L_{\rm ph}$} & \colhead{$L_{\rm d}$} & \colhead{$R_{\rm bb,d}$} \\
 		(days) & ($\rm 10^{42}\ erg\ s^{-1}$)  & ($\rm 10^{42}\ erg\ s^{-1}$) & ($\rm 10^{15}\ cm$) \\\hline
 		\multicolumn{4}{c}{\textbf{ASASSN-14li}}\\\hline
 		29.0 d & $34.82^{+2.89}_{-2.50}$ ($34.83^{+2.84}_{-2.47}$) & $0.37^{+0.23}_{-0.11}$ ($0.40^{+0.16}_{-0.09}$) & $11.09^{+3.53}_{-2.32}$ ($10.23^{+3.82}_{-2.07}$) \\\hline
 		199.0 d & $14.40^{+20.54}_{-7.54}$ ($14.33^{+21.77}_{-7.48}$) & $0.12^{+0.08}_{-0.03}$ ($0.15^{+0.06}_{-0.03}$) & $9.84^{+5.73}_{-3.22}$ ($9.56^{+7.17}_{-3.36}$) \\\hline
 		387.0 d & $6.08^{+12.21}_{-2.74}$ ($5.99^{+12.01}_{-2.67}$) & $0.21^{+0.12}_{-0.06}$ ($0.23^{+0.09}_{-0.05}$) & $9.01^{+3.38}_{-1.97}$ ($8.44^{+3.81}_{-1.88}$) \\\hline
	
 		\multicolumn{4}{c}{\textbf{ASASSN-15lh}}\\\hline
 		112.0 d & $510.19^{+48.09}_{-36.57}$ ($509.97^{+48.10}_{-36.52}$) & $18.91^{+6.76}_{-3.24}$ ($21.71^{+4.78}_{-3.00}$)  & $84.94^{+35.83}_{-22.15}$ ($80.10^{+37.08}_{-20.42}$) \\\hline
 		265.0 d & $156.06^{+7.52}_{-6.31}$ ($156.06^{+7.54}_{-6.33}$)  & $22.30^{+6.70}_{-3.13}$ ($26.63^{+4.99}_{-3.29}$)  & $108.30^{+43.57}_{-31.56}$ ($100.80^{+47.93}_{-29.35}$) \\\hline
 		405.0 d & $55.70^{+3.54}_{-3.38}$ ($55.73^{+3.66}_{-3.42}$)  & $24.47^{+2.16}_{-1.95}$ ($30.87^{+3.03}_{-2.65}$)  & $171.59^{+44.47}_{-38.00}$ ($166.84^{+53.58}_{-44.27}$) \\\hline

 		\multicolumn{4}{c}{\textbf{ASASSN-18ul}}\\\hline
 		
 		95.0 d & $239.63^{+71.69}_{-43.75}$ ($238.61^{+70.66}_{-43.51}$) & $0.53^{+0.34}_{-0.24}$ ($0.54^{+0.28}_{-0.22}$)  & $11.81^{+5.61}_{-2.74}$ ($11.47^{+6.77}_{-2.87}$) \\\hline
 		
 		\multicolumn{4}{c}{\textbf{ASASSN-18zj}}\\\hline
 		21.0 d & $95.82^{+0.99}_{-0.98}$ ($95.84^{+0.98}_{-0.98}$)  & $0.88^{+0.11}_{-0.08}$ ($1.17^{+0.12}_{-0.10}$)  & $37.69^{+10.31}_{-9.02}$ ($37.89^{+13.55}_{-10.98}$) \\\hline
 		170.0 d & $15.21^{+1.65}_{-1.44}$ ($15.21^{+1.66}_{-1.44}$)  & $0.32^{+0.21}_{-0.11}$ ($0.36^{+0.15}_{-0.10}$)  & $11.13^{+7.40}_{-2.84}$ ($11.02^{+8.56}_{-3.08}$) \\\hline

 		\multicolumn{4}{c}{\textbf{PS18kh}}\\\hline
 		17.0 d & $86.80^{+21.84}_{-15.13}$ ($87.38^{+22.21}_{-15.22}$)  & $0.54^{+0.29}_{-0.21}$ ($0.73^{+0.50}_{-0.29}$)  & $28.17^{+48.01}_{-16.13}$ ($29.49^{+68.57}_{-17.73}$) \\\hline

 		\multicolumn{4}{c}{\textbf{ZTF18acaqdaa}}\\\hline
 		91.0 d & $674.21^{+4247.29}_{-482.34}$ ($642.27^{+4065.02}_{-457.19}$)  & $6.61^{+3.21}_{-2.61}$ ($6.87^{+2.66}_{-2.54}$)  & $38.00^{+16.62}_{-8.37}$ ($37.17^{+18.92}_{-8.77}$) \\\hline

 	\end{tabular}
 \end{table*}

\clearpage

\clearpage

%
%

\clearpage

\clearpage

\clearpage

\begin{thebibliography}
	
\bibitem[Dong(2016)]{Dong2016}Dong, S., Shappee, B. J., Prieto, J. L., et al. 2016, Science, 351, 6270

\bibitem[Dou et al.(2016)]{Dou2016}Dou, L., Wang, T., Jiang, N., et al. 2016, \apj, 832, 188
\bibitem[Dou et al.(2017)]{Dou2017}Dou, L., Wang, T., Yan, L., et al. 2017, \apjl, 841, L8

\bibitem[Foreman-Mackey et al.(2013)]{Foreman-Mackey2013}Foreman-Mackey, D., Hogg, D. W., Lang, D., et al. 2013, \pasp, 125, 306

\bibitem[Fox et al.(2011)]{Fox2011}Fox, O., Chevalier, R., Skrutskie, M., et al. 2011, \apj, 741, 7

\bibitem[Frederick et al.(2019)]{Frederick2019}Frederick, S., Gezari, S., Graham, M. J., et al. 2019, \apj, 883, 31

\bibitem[Gall et al.(2014)]{Gall2014}Gall, C., Hjorth, J., Watson, D., et al. 2014, \nat, 511, 326

\bibitem[Gan et al.(2021)] {Gan2021}Gan, W., Wang, S., \& Liang, E.  2021, \apj, 914, 125

\bibitem[Gezari(2021)]{Gezari2021}Gezari, S. 2021, 2021, \araa, 59, 21

\bibitem[Godoy-Rivera et al.(2017)]{Godoy2017}Godoy-Rivera,D., Stanek,K., Kochanek,C., et al. 2017, \mnras, 466,2

\bibitem[Gomez et al.(2020)]{Gomez2020}Gomez, S., Nicholl, M., Philip, S., et al. 2020, \mnras, 497, 1925

\bibitem[Hills(1975)]{Hills1975}Hills, J. 1975, \nat, 254, 5498, 295

\bibitem[Holoien et al.(2016)]{Holoien2016}Holoien, T., Kochanek, C., Prieto, J., et al. 2016,\mnras, 455, 2918

\bibitem[Holoien(2019)]{Holoien2019} Holoien, T., Huber, M., et al. 2019, \apj, 880, 120

\bibitem[Jiang et al.(2016)]{Jiang2016}Jiang, N., Dou, L., Wang, T., et al. 2016, \apjl, 828, L14

\bibitem[Jiang et al.(2021)]{Jiang2021}Jiang, N., Wang, T., Hu, X., Sun, L., Dou, L., \& Xiao, L. 2021, \apj, 911, 31

\bibitem[Jiang et al.(2019)]{Jiang2019}Jiang, N., Wang, T., Mou, G., et al. 2019, \apj, 871, 15

\bibitem[Jiang et al.(2017)]{Jiang2017}Jiang, N., Wang, T., Yan, L., et al. 2017, \apj, 850, 63

\bibitem[Leloudas (2016)]{Leloudas2016}Leloudas, G., Fraser, M., Stone, N. C., et al.2016, NatAs, 1, 0002

\bibitem[Mattila et al.(2008)]{Mattila2008}Mattila, S., Meikle, W., Lundqvistet, P., et al. 2008, \mnras, 389, 141

\bibitem[Rees(1988)]{Rees1988}Rees, M. J. 1988, \nat, 333, 6173, 523

\bibitem[Stritzinger et al.(2012)]{Stritzinger2012}Stritzinger, M., Taddia, F., Fransson, C., et al. 2012, \apj, 756, 173

\bibitem[van Velzen et al.(2016)]{van Velzen2016}van Velzen, S., Mendez, A., Krolik, J., \& Gorjian, V. 2016, \apj, 829, 19

\bibitem[van Velzen et al.(2020)]{van Velzen2020}van Velzen, S., Holoien, T., Onori, F., et al. 2020, SSRv, 216, 124

\bibitem[Wevers, T. (2019)]{Wevers2019} Wevers, T., Pasham, D., van Velzen, S., et al. 2019, \mnras, 488,4816

\end{thebibliography}
\end{document}